\documentclass[iop,revtex4]{emulateapj}

\usepackage{apjfonts}
\usepackage{natbib}
\usepackage{amssymb}
\usepackage{enumitem}
\usepackage{longtable}
\usepackage[breaklinks,colorlinks,citecolor=blue,linkcolor=magenta]{hyperref}
\usepackage{lscape}
\usepackage{multirow}

\usepackage{color}

\newcommand{\sysanal}{\hbox{{\sc sysanal}}}
\newcommand{\minfit}{\hbox{{\sc minfit}}}
\newcommand{\vpfit}{\hbox{\sc vpfit}}
\newcommand{\magiicat}{\hbox{{\rm MAG}{\sc ii}CAT}}
\newcommand{\MgIIdblt}{{\rm Mg}\kern 0.1em{\sc ii}~$\lambda\lambda 2796, 2803$}
\newcommand{\MgII}{\hbox{{\rm Mg}\kern 0.1em{\sc ii}}}
\newcommand{\OVIdblt}{{\rm O}\kern 0.1em{\sc vi}~$\lambda\lambda 1031, 1037$} 
\newcommand{\OVI}{\hbox{{\rm O}\kern 0.1em{\sc vi}}}
\newcommand{\CIV}{\hbox{{\rm C}\kern 0.1em{\sc iv}}}
\newcommand{\HI}{\hbox{{\rm H}\kern 0.1em{\sc i}}}
\newcommand{\Lya}{\hbox{{\rm Ly}\kern 0.1em$\alpha$}}
\newcommand{\Lyb}{\hbox{{\rm Ly}\kern 0.1em$\beta$}}
\newcommand{\OII}{\hbox{{\rm O}\kern 0.1em{\sc ii}}}
\newcommand{\OIII}{\hbox{{\rm O}\kern 0.1em{\sc iii}}}
\newcommand{\NII}{\hbox{{\rm N}\kern 0.1em{\sc ii}}}
\newcommand{\kms}{\hbox{km~s$^{-1}$}}
\newcommand{\etal}{et~al.}

\newcommand{\vfifty}{\hbox{$\Delta v(50)$}}
\newcommand{\vninety}{\hbox{$\Delta v(90)$}}

\shorttitle{~{\MgII} Intragroup Medium Kinematics}
\shortauthors{\sc Nielsen {\etal}}
\slugcomment{ApJ accepted 31 October 2018}

\begin{document}

\title{~{\magiicat} VI. The {\MgII} Intragroup Medium is Kinematically Complex}

\author{
Nikole M. Nielsen$^{1}$, 
Glenn G. Kacprzak$^{1}$,
Stephanie K. Pointon$^{1}$, 
Christopher W. Churchill$^{2}$, \\
and 
Michael T. Murphy$^{1}$
}

\affil{$^1$ Centre for Astrophysics and Supercomputing, Swinburne
  University of Technology, Hawthorn, Victoria 3122, Australia;
  nikolenielsen@swin.edu.au\\
  $^2$ Department of Astronomy, New Mexico State University, Las
  Cruces, NM 88003, USA
}

\begin{abstract}

By comparing {\MgII} absorption in the circumgalactic medium (CGM) of
group environments to isolated galaxies, we investigated the impact of
environment on the CGM. A {\MgII} absorber is associated with a group
if there are two or more galaxies at the absorption redshift within a
projected distance of $D=200$~kpc from a background quasar and a
line-of-sight velocity separation of 500~{\kms}. We compiled a sample
of 29 group environments consisting of 74 galaxies ($2-5$ galaxies per
group) at $0.113<z_{\rm gal}<0.888$. The group absorber median
equivalent width ($\langle W_r(2796)\rangle=0.65\pm0.13$~{\AA}) and
covering fraction ($f_c=0.89_{-0.09}^{+0.05}$) are larger than
isolated absorbers ($1.27\sigma$ and $2.2\sigma$, respectively) but
median column densities are statistically consistent. A pixel-velocity
two-point correlation function analysis shows that group environment
kinematics are statistically comparable to isolated environments
($0.8\sigma$), but with more power for high velocity dispersions
similar to outflow kinematics. Group absorbers display more optical
depth at larger velocities. A superposition model in which multiple
galaxies contribute to the observed gas matches larger equivalent
width group absorbers, but overpredicts the kinematics significantly
due to large velocity separations between member galaxies. Finally,
galaxy--galaxy groups (similar member galaxy luminosities) may have
larger absorber median equivalent widths ($1.7\sigma$) and velocity
dispersions ($2.5\sigma$) than galaxy--dwarf groups (disparate
luminosities). We suggest the observed gas is coupled to the group
rather than individual galaxies, forming an intragroup medium. Gas may
be deposited into this medium by multiple galaxies via outflowing
winds undergoing an intergalactic transfer between member galaxies or
from tidal stripping of interacting members.

\end{abstract}


\section{Introduction}
\label{sec:intro}

Extensive work has gone into investigating the role that the baryon
cycle plays in forming galaxies and steering their evolution, with
particular focus on gas reservoirs such as the circumgalactic medium
(CGM). It is well-known that the baryon cycle regulates star formation
in galaxies via a balance of inflowing and outflowing gas
\citep[e.g.,][]{oppenheimer08, lilly-bathtub}, processes which must
take place in and contribute material to the CGM of galaxies. The
build-up of material into the CGM results in a gas reservoir with a
mass comparable to the interstellar medium \citep[][]{thom11,
  tumlinson11, werk13, peeples14} out to large distances
\citep[$D\gtrsim150$~kpc; e.g.,][and references therein]{chen10a,
  tumlinson11, rudie12, magiicat2}. Thus, the CGM represents an
excellent laboratory for studying the processes which control galaxy
evolution, containing remnants of past evolutionary processes and the
fuel for future star formation.

Using background quasar sightlines probing gas traced by the
{\MgIIdblt} absorption doublet (and other ion tracers), we now have a
simple picture of the CGM in which gas accretes onto galaxies along
their major axis to feed the ISM for future star formation
\citep[e.g.,][]{steidel02, ggk-sims, kcn12, stewart11, danovich12,
  danovich15, martin12, rubin-accretion, bouche13} and gas outflows
along the minor axis to further pollute the CGM with metal-enriched
gas \citep[e.g.,][]{rubin-winds, rubin-winds14, bouche12, kcn12,
  martin12, bordoloi14, bordoloi14-model, kacprzak14,
  schroetter16}. However, the large majority of this body of work has
focused on an environment in which only a fraction of galaxies are
found: isolated environments. Absorbers associated with groups and
clusters of galaxies have often been neglected and largely removed
from the analyses.

Galaxy evolution is also environment-dependent. Even before the most
complex parts of mergers occur, the signatures of galaxy--galaxy
interactions are observable. Observations of cool {\HI} gas show a
variety of structures due to galaxy interactions in group
environments, including tidal streams and filaments, warped disks, and
high velocity clouds \citep[e.g.,][]{fraternali02, chynoweth08,
  sancisi08, mihos12, wolfe13}. Using the Illustris simulations,
\citet{hani18} studied the impact of a major merger on the
circumgalactic medium and found that the covering fraction of the
largest column density gas increases pre-merger and remains elevated
for several billion years post-merger. This effect was due to
merger-driven outflows rather than tidal stripping. In the FIRE
simulations, \citet{angles17} also found that intergalactic transfer,
particularly the transfer of gas from the outflows of one galaxy onto
another nearby galaxy, is a dominant accretion mechanism of galaxies
by redshift $z=0$. These structures and the hierarchical processes
that place them between galaxies are an additional level of complexity
on top of the isolated galaxy CGM, yet understanding the CGM in these
denser environments is necessary for understanding how galaxies grow
and evolve. Just as the visible (emitting) portions of galaxies become
tidally stripped and disturbed, so should the diffuse (absorbing)
material in the CGM undergo complex interactions, and may do so before
the visible galaxy due to the large radii involved.

In cluster environments, \citet{lopez08} studied {\MgII} and found an
overabundance of strong {\MgII} absorbers that is more pronounced at
lower impact parameters, suggesting that the halos of cluster galaxies
are truncated at 10~kpc \citep[also see][]{padilla09, andrews13}. The
authors also found a relative lack of weak absorbers, which are
expected to be more easily destroyed in clusters where the numbers are
more consistent with those associated with isolated galaxies. Also on
an extreme end are ``ultrastrong'' {\MgII} absorbers with
$W_r(2796)\geqslant3$~{\AA}. Without determining galaxy redshifts,
\citet{nestor07} found evidence for a significant excess of galaxies
around quasar sightlines hosting these absorbers compared to random
fields, suggesting that group environments may give rise to some
fraction of these extreme absorbers in addition to starbursts and very
low impact parameter galaxies. Of the three ultrastrong {\MgII}
absorbers for which galaxy redshifts have been spectroscopically
determined \citep{nestor11, gauthier13}, all were found to be located
in group environments and interpreted to be either outflows as the
result of interaction-induced star formation, or tidal stripping.

In group environments, of which several have been studied,
\citet{chen10a} found that the equivalent widths of {\MgII} absorbers
in groups were similar to those associated with isolated galaxies, but
they did not exhibit an anti-correlation between equivalent width and
impact parameter, which has long been known for isolated galaxies
\citep[e.g.,][]{lanzetta90, sdp94, ggk08, chen10a, magiicat2}. Using
stacked galaxy spectra probing foreground galaxies, \citet{bordoloi11}
found that {\MgII} is more extended around groups, and this could be
explained by a superposition of the equivalent widths of member group
galaxies. Because of this superposition model, the authors suggest
that the group environment (i.e., tidal stripping, interaction-induced
star formation-driven outflows) does not appear to change the
properties of {\MgII} absorbers for individual galaxies. Finally,
\citet{whiting06}, \citet{ggk1127}, \citet{bielby17}, and
\citet{peroux17} studied the absorption in one or two group
environments each and concluded the gas was due to an intragroup
medium or tidal interactions depending on the detailed characteristics
of the sample. However, \citet{rahmani18} attributed the observed
absorption to a single galaxy in the group, partially from the stellar
disk and partially accretion onto a warped disk.

We focus on a sample of group galaxies compiled during our work to
form the {\MgII} Absorber--Galaxy Catalog
\citep[{\magiicat};][]{magiicat2, magiicat1, magiicat5, magiicat4,
  magiicat3}. Because of this, we did not actively seek out galaxies
obviously undergoing mergers/interactions and therefore, the galaxies
presented here are likely pre-merger but are still expected to show
the effects of residing in more dense environments. While the galaxies
themselves may not be obviously merging, their CGM is likely already
affected by the group environment due to the large radius of the CGM
out to roughly 200~kpc, compared to the visible (in emission) portions
of the galaxies.

The paper is organized as follows: Section~\ref{sec:methods} describes
our galaxy and quasar samples, along with our methods for creating a
standardized catalog of group absorber--galaxy
pairs. Section~\ref{sec:EWD} details the properties of the group
sample compared to the isolated {\magiicat} sample for the
anti-correlation between {\MgII} equivalent width and impact parameter
while Section~\ref{sec:kinematics} examines the absorption kinematics
with the pixel-velocity two-point correlation function. These sections
also report the results of a superposition model in which multiple
galaxies contribute to the CGM of group galaxies. We examine the
absorber Voigt profile cloud column densities and velocities in
Section~\ref{sec:NvsV}. Section~\ref{sec:discussion} discusses the
impact of the group environment on the CGM. Finally,
Section~\ref{sec:conclusions} summarizes the work. We adopt a
$\Lambda$CDM cosmology ($H_0=70$ km s$^{-1}$ Mpc$^{-1}$,
$\Omega_M=0.3$, and $\Omega_{\Lambda}=0.7$) and report AB absolute
magnitudes throughout this paper. The group catalog presented here has
been placed on-line at the NMSU Quasar Absorption Line Group
website\footnote{http://astronomy.nmsu.edu/cwc/Group/magiicat} along
with the previously published isolated galaxy sample.

\begin{figure*}[ht]
  \includegraphics[width=\linewidth]{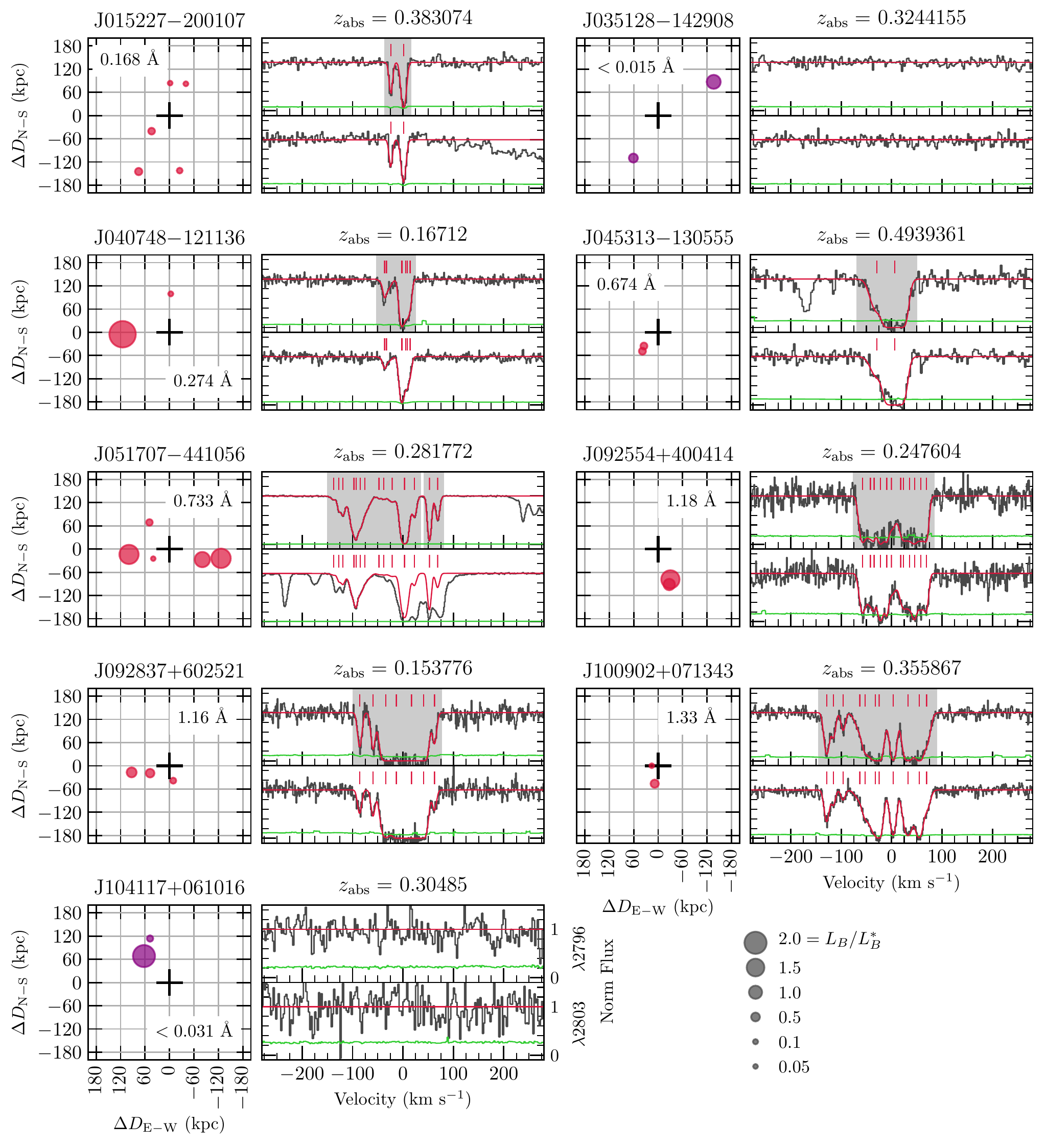}
  \caption[]{On-the-sky locations and absorption spectra for each
    group environment with measured {\MgII} absorption and a
    high-resolution HIRES/Keck or UVES/VLT spectrum. The left panel
    for each group shows the locations of each group galaxy (red and
    purple points) in physical space relative to the associated
    background quasar (black cross). Point sizes represent galaxy
    luminosity, $L_B/L_B^{\ast}$, with larger points representing more
    luminous galaxies. Red points represent those absorbers used in
    our kinematics analysis, while purple are not included in the
    kinematics analysis. The top panel in each spectrum panel pair
    shows the {\MgII}~$\lambda 2796$ line, while the bottom panel
    shows the {\MgII}~$\lambda 2803$ line. Black histograms are the
    data, red curves are the fit to the spectrum, red ticks are the
    individual Voigt profile components, and the green data are the
    error spectrum. Regions of the spectra where we use the pixel
    velocities for our kinematic analysis are highlighted in gray. The
    velocity zero points are determined by the optical depth-weighted
    median of absorption. Measured $W_r(2796)$ values are listed in
    the left panels for each group. We only have an upper limit on
    absorption for the J035128$-$142908 (Q0349$-$146) and
    J104117$+$061016 (Q1038$+$064) fields, and so there are no gray
    shaded regions.}
  \label{fig:radecspec}
\end{figure*}
\addtocounter{figure}{-1}

\begin{figure*}[ht]
  \includegraphics[width=\linewidth]{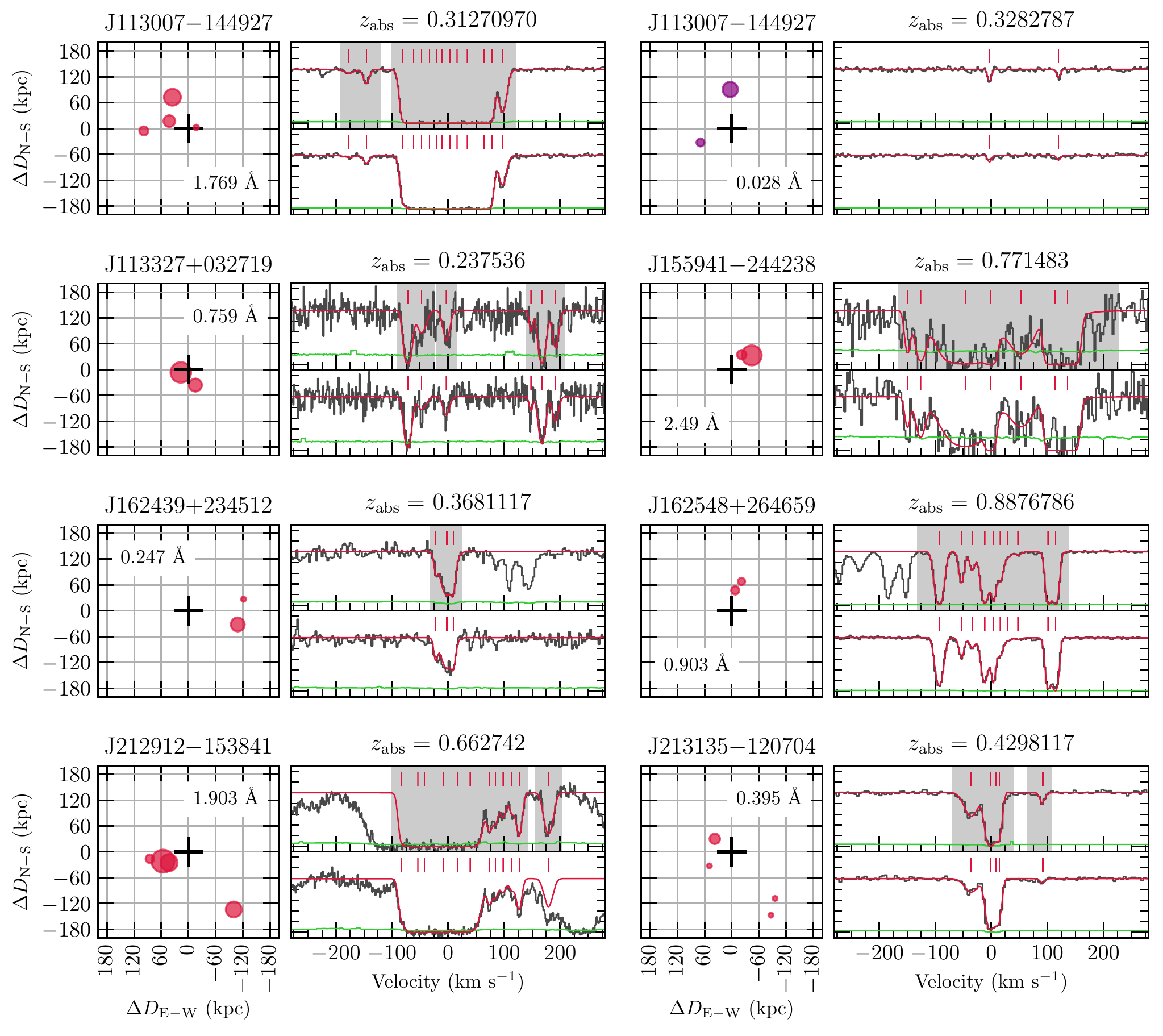}
  \caption[]{(continued) The absorber in J113007$-$144927
    (Q1127$-$145), $z_{\rm gal}=0.328$ does not have gray shaded
    regions because the equivalent width of this absorber is below our
    equivalent width detection threshold, which we applied to ensure a
    uniform kinematic sample.}
  \label{fig:radecspec2}
\end{figure*}

\begin{figure*}[ht]
  \centering
  \includegraphics[scale=0.77]{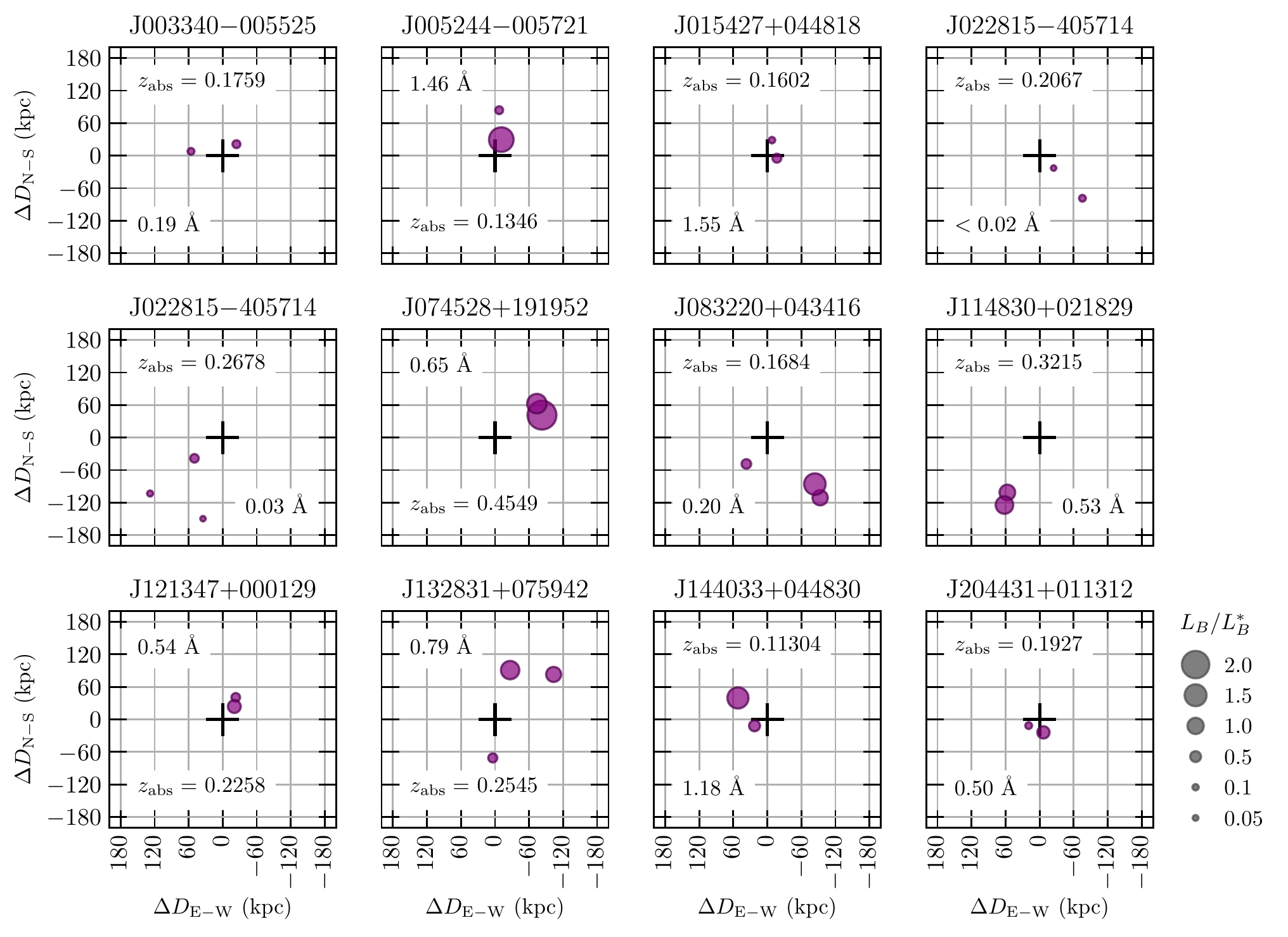}
  \caption[]{On-the-sky locations of each group galaxy in physical
    space for those groups in which we do not have high-resolution
    spectra of the associated background quasar. However, equivalent
    widths were measured for each associated absorber and are listed
    in Table~\ref{tab:calcprops} (including the measurement source),
    as well as on each panel. Purple points represent each galaxy in
    the group and the black cross represents the associated background
    quasar. Point sizes represent galaxy luminosity, $L_B/L_B^{\ast}$,
    with larger points representing more luminous galaxies. Galaxies
    in this Figure are not included in the kinematic TPCF analysis,
    but are included in Figures~\ref{fig:EWD}-\ref{fig:superboot}.}
  \label{fig:nospecradec}
\end{figure*}

\section{Data and Methods}
\label{sec:methods}

We compiled a sample of 29 {\MgII} absorbers along 27 quasar
sightlines and associated with a total of 74 foreground galaxies in
group environments. The galaxies are located at $0.113 < z_{\rm gal} <
0.888$ and within a projected distance of $D=200$~kpc from the
background quasar. An absorber is classified as being located in a
group environment if there are two or more associated galaxies within
a projected distance of 200~kpc and the galaxies have a line-of-sight
velocity separation of less than 500~{\kms}. See \citet[][hereafter
  {\magiicat} I]{magiicat1} for further discussion of our group
environment criteria. While it is not one of the selection criteria, a
majority of the groups in the sample are close ($\lesssim50$~kpc)
pairs of galaxies with similar luminosities. Galaxy luminosities have
a range of $0.01 < L_B/L_B^{\ast} < 2.49$ for all group galaxies or
$0.15 < L_B/L_B^{\ast} < 2.49$ for only the most luminous galaxy in a
group. Galaxy luminosity ratios (most luminous to second-most
luminous) range from $1.01 < L_1/L_2 < 56.0$, where most have ratios
below $L_1/L_2 = 10$.

In the following sections, we further describe the group sample and
the sources of the data. We also describe the quasar spectra and their
analysis.

\subsection{Group Galaxy Sample}
\label{sec:groups}

The group sample presented here was largely identified during our work
to create the isolated galaxy sample in the {\MgII} Absorber--Galaxy
Catalog ({\magiicat}) where we either cataloged galaxies already
identified as groups in the literature, or we identified new groups
when consolidating multiple sources of data in the same fields. These
galaxies are sourced from \citet{sdp94}, Steidel (1996, private
communication), \citet{guillemin97}, \citet{steidel97},
\citet{chen10a}, \citet{ggk1127}, \citet{kacprzak11kin}, and
\citet{kcems11}. The surveys presented in each of these are detailed
in {\magiicat} I. We obtained the published galaxy data for several
more group environments from \citet{whiting06}, \citet{bielby17},
\citet{peroux17}, \citet{pointon17}, and \citet{rahmani18}, and
include new data for the Q1038$+$064 field, all of which we describe
below. To summarize, the survey methods for these literature sources
include absorption-selected samples, gas cross-section-selected
samples (within a given impact parameter expected for {\MgII} halos),
``control fields'' that were purposely targeted because absorption was
not detected, magnitude-limited samples, and volume-limited samples. 

There are additional groups published in \citet{nestor11} and
\citet{gauthier13}, though they are classified as ``ultrastrong''
      {\MgII} absorbers ($W_r(2796)\geq 3$~{\AA}). Due to their large
      equivalent widths and kinematic spreads, we therefore consider
      these absorbers outliers compared to the rest of our sample
      described below and this is further discussed in
      Section~\ref{sec:discussion}. We refrain from including these
      absorber--galaxy pairs in this sample and we also exclude the
      single isolated ultrastrong {\MgII} absorber from the isolated
      galaxy analyses.

\subsubsection{\citet{whiting06}}

Working with the known $z_{\rm abs}=0.663$ {\MgII} absorber in the PKS
2126$-$158 field (J212912$-$153841), \citet{whiting06}
identified a group of galaxies at the redshift of absorption. The
authors observed the field with the GMOS multi-object spectroscopy
mode on Gemini South and imaged in the $i'$ band. Galaxies were
observed out to a field of view of $\sim 5'.5$ and down to a limiting
magnitude of $i'=24.6$. Eight galaxies were observed at $z\sim 0.66$,
but only five were located within $D=200$~kpc of the quasar sightline,
and the redshift of one of the five galaxies is larger than our
line-of-sight velocity separation criterion to be considered a group
galaxy. We remeasured the equivalent width of this absorber in a
UVES/VLT spectrum of the background quasar.

\subsubsection{\citet{bielby17}}

Observing with the Multi Unit Spectroscopic Explorer (MUSE) on the
VLT, \citet{bielby17} spectroscopically identified a group of five
galaxies in the HE0515$-$4414 (J051707$-$441056) field at the redshift
of a $z=0.282$ {\MgII} absorber. Galaxy apparent magnitudes were
calculated in the $R$-band and the MUSE data cube has a $3\sigma$
depth of $f=16\times10^{-18}$~erg~cm$^{-2}$~s$^{-1}$~\AA$^{-1}$. We
obtained the UVES/VLT high signal-to-noise spectrum \citep{kotus17}
and modeled the absorber following the methods described in
Section~\ref{sec:qso} to be consistent with our previous work.

\newpage
\subsubsection{\citet{peroux17}}

\citet{peroux17} observed the $z_{\rm abs}=0.4298$ absorber in the
Q2128$-$123 field (J213135$-$120704) with MUSE/VLT to investigate the
environment of the previously known absorber and its assumed isolated
galaxy host. From two pointings with exposure times of 1200 s, the
authors found an additional three low-luminosity
($L/L^{\ast}\sim0.01$) galaxies at the redshift of the absorber. This
field was classified as an isolated pair in {\magiicat} I, but is now
included in the present sample with the new findings. We remeasured
the magnitudes of the two brightest galaxies in the group from a
WFPC2/{\it HST} F207W image, but adopt the magnitudes and luminosities
for the two faintest galaxies from \citet{peroux17} due to their being
too faint to detect in the {\it HST} image.

\subsubsection{\citet{pointon17}}

The groups compiled by \citet{pointon17} were selected for having
COS/{\it HST} spectra that covered the wavelength at which {\OVI}
absorption due to group environments was expected. From their sample,
we selected groups for which HIRES/Keck and/or UVES/VLT spectra
covered the {\MgII} doublet, regardless of whether absorption was
detected, and measured the {\MgII} equivalent width or a $3\sigma$
upper limit on $W_r(2796)$. We also enforced the impact parameter and
galaxy--galaxy velocity separation criteria for {\MgII} groups
described in Section~\ref{sec:methods}, which is more constraining
than the {\OVI} group criterion published by \citet{pointon17}. The
galaxies drawn from this work were originally published in
\citet{chen01b}, \citet{chen09}, \citet{meiring11}, \citet{werk12},
and \citet{johnson13}. From these works, we found three absorbers that
were initially classified as isolated absorber--galaxy pairs in
{\magiicat} I, but have moved them to the group sample. These include
the fields J022815$-$405714 ($z_{\rm abs} = 0.2067, 0.2678$) and
J035128$-$142908 ($z_{\rm abs}=0.3244$).

\subsubsection{\citet{rahmani18}}

Observing another previously known {\MgII} absorber assumed to be
associated with an isolated galaxy (Q0150$-$202, J015227$-$200107,
$z_{\rm abs}=0.383$), \citet{rahmani18} found an additional five
galaxies with spectroscopic redshifts at the absorber redshift. The
authors imaged the field with MUSE/VLT for a total of 100 min across
two exposures, covering galaxies out to impact parameters of
$\sim200$~kpc. As already stated, this absorber--galaxy pair was
previously identified as isolated in {\magiicat} I, but we have moved
the field to the present sample. Finally, we remeasured the galaxy
magnitudes from a WFPC2/{\it HST} F702W image to be consistent with
our measurements of the assumed isolated host.

\subsubsection{Field Q1038$+$064}

The $z_{\rm gal}=0.3044$ galaxy in this field (also known as
J104117$+$061016) was identified, and its properties and associated
quasar spectrum were provided to us by C.~Steidel (1996, private
communication). We obtained the spectrum and spectroscopic redshift of
the $z_{\rm gal}=0.3053$ galaxy with the Dual Imaging Spectrograph
(DIS) on the Apache Point Observatory 3.5m telescope in March 2008 and
the data were reduced using standard methods using {\sc
  iraf}.\footnote{{\sc iraf} is distributed by the National Optical
  Astronomy Observatory, which is operated by the Association of
  Universities for Research in Astronomy under cooperative agreement
  with the National Science Foundation.} This is one of only three
group fields in the sample presented here to have only an upper limit
on {\MgII} absorption measured.

\subsubsection{Galaxy Properties}

Details of the methods used to determine galaxy properties are
described in full in {\magiicat} I (Section 3.1 and the Appendices),
as we compiled the majority of the group sample with the isolated
sample. The galaxy properties obtained from the new group sample
publications listed above were converted to AB $B$-band absolute
magnitudes and luminosities and the $\Lambda$CDM cosmology ($H_0=70$
km s$^{-1}$ Mpc$^{-1}$, $\Omega_M=0.3$, and $\Omega_{\Lambda}=0.7$)
following the methods presented in {\magiicat} I.

We obtained new galaxy spectra in eight fields (14 galaxies) with the
Keck Echelle Spectrograph and Imager
\citep[ESI;][]{sheinis02}. Details of the data reduction are presented
in \citet{kacprzak18}, but the aim was to obtain accurate galaxy
redshifts with precisions of $3-20$~{\kms}. The ESI spectra have a
resolution of 22~{\kms}~pixel$^{-1}$ when binned by two and cover a
wavelength range of 4000 to 10,000~{\AA}. Emission lines covered in
this range include the {\OII} doublet, $\rm{H}\beta$, the {\OIII}
doublet, $\rm{H}\alpha$, and the {\NII} doublet. Galaxy spectra were
vacuum and heliocentric velocity corrected for direct comparison with
the absorption line spectra. Finally, the Gaussian fitting algorithm
\citep[FITTER; see][]{archiveI} was used to determine the best-fit
centroids and widths of the covered emission lines to determine galaxy
redshifts.

Observed galaxy properties are tabulated in
Table~\ref{tab:obsprops}. The columns are: (1) QSO identifier, (2),
Julian 2000 designation (J-Name), (3) galaxy spectroscopic redshift,
$z_{\rm gal}$, (4) quasar--galaxy right ascension offset, $\Delta
\alpha$, (5) quasar--galaxy declination offset, $\Delta \delta$, (6)
quasar--galaxy angular separation, $\theta$, (7) reference for Columns
4, 5, and 6, (8) apparent magnitude used to obtain $M_B$, (9) band for
the preceding apparent magnitude, (10) reference for Columns 8 and 9,
(11) apparent magnitude used to calculate $M_K$, (12) band for $m_K$,
(13) reference for Columns 11 and 12, and (14) galaxy SED type
\citep[from][]{cww80, bolzonella00} based on the galaxy observed
color.

Calculated galaxy properties are tabulated in
Table~\ref{tab:calcprops}. Columns are: (1) QSO identifier, (2) Julian
2000 designation (J-Name), (3) galaxy spectroscopic redshift, $z_{\rm
  gal}$, (4) {\MgII} absorption redshift, $z_{\rm abs}$, (5) {\MgII}
rest equivalent width, $W_r(2796)$, (6) {\MgII} doublet ratio, (7)
reference for Columns 4, 5, and 6, (8) quasar--galaxy impact
parameter, $D$, (9) $K$-correction to obtain $M_B$, (10) absolute
$B$-band magnitude, $M_B$, (11) $B$-band luminosity, $L_B/L_B^{\ast}$,
(12) $K$-correction to obtain $M_K$, (13) absolute $K$-band magnitude,
$M_K$, (14) $K$-band luminosity, $L_K/L_K^{\ast}$, and (15) rest-frame
color, $B-K$.

To illustrate their positions relative to each other and the quasar
sightline, galaxies are plotted in RA and Dec (with physical
distances) from the background quasar sightline (cross) in
Figures~\ref{fig:radecspec} (square panels) and
\ref{fig:nospecradec}. Point sizes represent galaxy $B$-band
luminosities, $L_B/L_B^{\ast}$, where larger points are more luminous
galaxies.

\subsection{Quasar Spectra}
\label{sec:qso}

We have high-resolution quasar spectra for 16 fields (17 group
environments) from HIRES on Keck or UVES on the VLT. Most of the
spectra have been published elsewhere \citep{cwc-thesis, cv01,
  evans-thesis, kcems11, werk13, kotus17}. The J155941$-$244238 quasar
was observed specifically for this work in March 2013 with UVES on the
VLT (programme number: 090.A-0304(A)) in the custom DIC2$-$470$+$760nm
setting for a total exposure time of 2660s. The spectrum was reduced
with the UVES pipeline \citep{dekker-uves} and the exposures were
combined and continuum fit with UVES\_popler \citep{uvespopler,
  murphy18}.

To obtain the CGM absorption properties from these high-resolution
spectra, the {\MgIIdblt} doublet absorption was modeled using one of
two methods: (1) a combination of {\sysanal} and {\minfit} for six
absorbers and (2) {\vpfit} for nine. The methods are summarized below.

The absorbers in the J045313$-$130555, J113007$-$144927,
J162439$+$234512, J162548$+$264659, and J213135$-$120704 fields were
modeled using {\sysanal} and {\minfit}, the methods for which are
detailed in \citet{cwc-thesis}, \citet{cv01}, \citet{cvc03}, and
\citet{evans-thesis}. {\sysanal} detects {\MgII} absorption with a
$5\sigma$ ($3\sigma$) significance criterion in the $\lambda 2796$
($\lambda 2803$) line following the formalism of
\citet{schneider93}. The code determines wavelength and velocity
bounds where absorption is formally detected and calculates the
rest-frame equivalent width, $W_r(2796)$. The absorption redshift,
$z_{\rm abs}$, is defined by the median wavelength of the apparent
optical depth distribution of absorption. All systems are then fit
using Voigt profile (VP) decomposition with {\minfit}
\citep{cwc-thesis, cv01, cvc03, evans-thesis} and the model with the
fewest statistically significant VP components (clouds) is
adopted. Cloud velocities, column densities, and Doppler $b$
parameters are obtained from the {\minfit} analysis.

For the remaining absorbers, J015227$-$200107, J040748$-$121136,
J051707$-$441056, J092554$+$400414, J092837$+$602521,
J100902$+$071343, J113327$+$032719, J155941$-$244238, and
J212912$-$153841, we used {\vpfit} \citep{vpfit}, and the full method
is described in \citet{pointon17}. Absorption redshifts are defined as
the optical depth-weighted median of absorption as above and the
velocity bounds of absorption were determined by finding the pixels at
which the VP model decreases by $1\%$ from the continuum level. The
two fitting methods are comparable and do not result in any
significant differences in our results.

The spectra and fits for each absorber are plotted in the second and
fourth columns of Figure~\ref{fig:radecspec} for the 17 absorbers for
which we have spectra. Black histograms are the data, red lines the
model, green lines the error spectrum, and red ticks are the
individual Voigt profile components. Shaded regions represent the
velocity range of absorption for the $\lambda 2796$ line. Panels
without shaded regions are either absorbers for which we have only a
$3\sigma$ upper limit on absorption or the absorber has an equivalent
width lower than the spectral equivalent width sensitivity limit of
$0.04$~{\AA} \citep[see][hereafter {\magiicat} IV]{magiicat4}.

In cases where HIRES/Keck and/or UVES/VLT spectra are not available,
we adopted the best published {\MgII} absorption values, typically the
most recent measurements or those obtained from the highest resolution
quasar spectra. These values and the references from which we obtained
the values are tabulated in Table~\ref{tab:calcprops}. Upper limits on
absorption are quoted at $3\sigma$.

\subsection{Isolated Galaxy Sample}

To test the influence that environment has on the CGM, we compare the
group sample described above to our previously published isolated
galaxy sample ({\magiicat} I). This sample has been modified to
reflect new information on environments as detailed in
Section~\ref{sec:groups} and to add the increasing number of
spectroscopically-confirmed {\MgII} absorber--galaxy pairs published
in the literature. Thus {\magiicat} is a living catalog and its
changes are periodically recorded on our publicly accessible
website.\footnote{http://astronomy.nmsu.edu/cwc/Group/magiicat}

\section{Equivalent Width vs Impact Parameter}
\label{sec:EWD}

Here we examine the anti-correlation between equivalent width and
impact parameter for the group galaxy sample described in the previous
section compared to our isolated galaxy sample from {\magiicat} I.

\subsection{$W_r(2796)$ vs. $D$: All Group Galaxies}

\begin{figure}[t]
  \includegraphics[width=\linewidth]{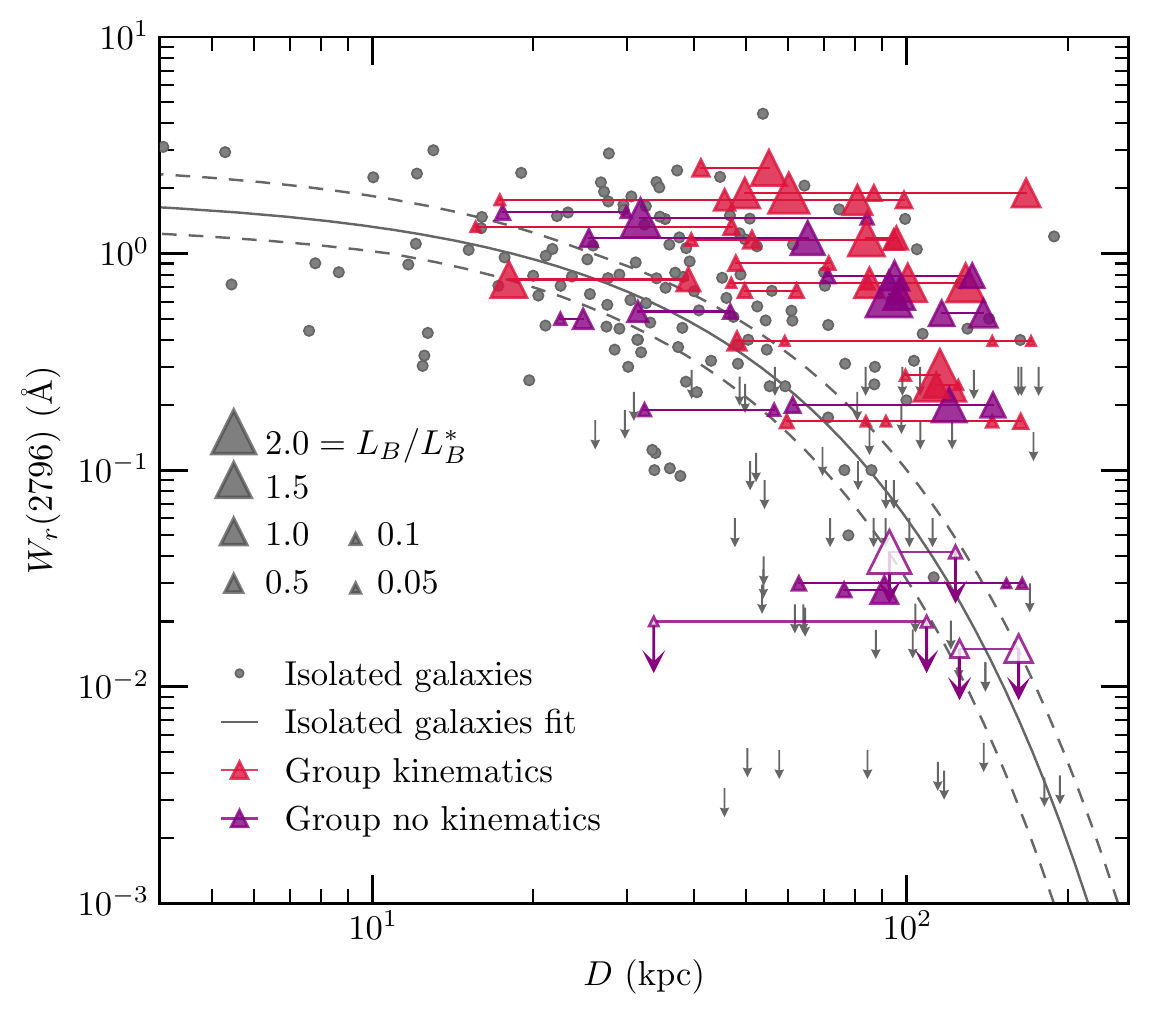}
  \caption[]{{\MgII} equivalent width, $W_r(2796)$, as a function of
    impact parameter, $D$. Gray points represent absorbers (solid
    points) and non-absorbers (downward arrows) associated with
    galaxies in isolated environments. Purple and red triangles
    represent absorbers associated with galaxies in group
    environments, where the point sizes represent their $B$-band
    luminosity, $L_B/L_B^{\ast}$. For each group, we plot every galaxy
    in the group at the equivalent width of the absorber with a
    horizontal line drawn through each galaxy. Red triangles are those
    absorbers included in our kinematics analysis, while purple
    triangles are those for which we do not have a high-resolution
    spectrum of the background quasar or the measured equivalent width
    (including limits) is lower than our completeness cut of
    0.04~{\AA} for the kinematics study.}
  \label{fig:EWD}
\end{figure}

A well-known relationship between the CGM and host galaxy properties
is the {\MgII} equivalent width anti-correlation with impact
parameter, $W_r(2796)$ vs $D$ \citep[e.g.,][]{lanzetta90, bb91,
  steidel95, chen10a, kcems11, magiicat2,
  magiicat1}. Figure~\ref{fig:EWD} presents this anti-correlation for
all group galaxies and the isolated galaxies from {\magiicat}~I and
\citet{magiicat2}, hereafter {\magiicat}~II. Gray points and downward
arrows correspond to the isolated galaxies and the solid and dashed
gray lines are the log-linear fit and uncertainties to the isolated
galaxy data from {\magiicat}~II. Because the group sample has multiple
galaxies associated with a single {\MgII} absorber, there are galaxies
at several impact parameters with the same $W_r(2796)$. The groups are
identified by triangle points connected by horizontal lines. Point
colors correspond to those in Figure~\ref{fig:radecspec}, where red
triangles are those groups for which we have high resolution quasar
spectra and a measured equivalent width above an equivalent width
completeness cut of 0.04~{\AA}. Purple points are the rest of the
group sample.

From Figure~\ref{fig:EWD} it appears that absorbers in group
environments have larger equivalent widths at a given impact parameter
than for the isolated sample. The median (mean) equivalent widths for
the group and isolated galaxy samples are $\langle W_r(2796)\rangle =
0.65\pm0.13$~{\AA} ($0.75\pm0.11$~{\AA}) and $\langle W_r(2796)\rangle
= 0.41\pm0.06$~{\AA} ($0.62\pm0.05$~{\AA}), respectively for the full
sample. Upper limits on the equivalent width were considered
``measurements'' at the upper limit value. The median equivalent
widths for the full group sample are larger than for the isolated
sample ($1.7\sigma$).

\begin{figure*}[ht]
  \centering
  \includegraphics[scale=0.75]{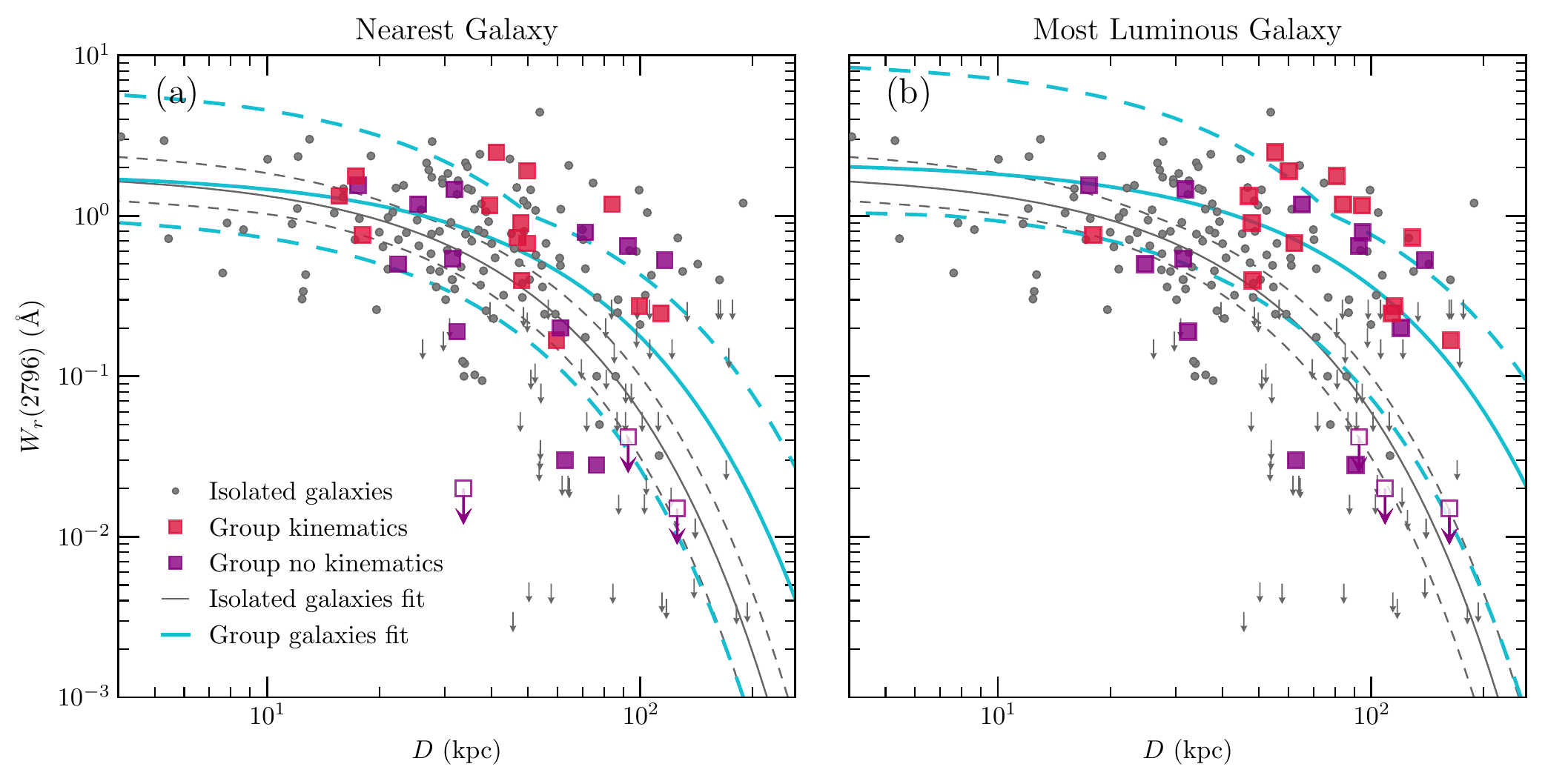}
  \caption[]{{\MgII} equivalent width, $W_r(2796)$, as a function of
    impact parameter, $D$, for (a) the galaxy nearest to the quasar
    sight-line and (b) the most luminous galaxy in the group. Isolated
    galaxies are represented by gray points and arrows, while group
    galaxies are squares. Square point colors are the same as the
    triangles in Figure~\ref{fig:EWD}. The gray solid and dashed
    curves are the expectation-maximization maximum-likelihood model
    and its $1\sigma$ uncertainties, respectively, published in
    \citet{magiicat2} for the isolated galaxy sample. Cyan solid and
    dashed curves are the model and $1\sigma$ uncertainties,
    respectively, for the full group sample (red and purple points)
    plotted in each panel. The CGM for group environments has a
    similar distribution to isolated galaxies on the $W_r(2796)-D$
    plane when the nearest galaxy is assumed to host the absorption,
    but the fitted slope may flatter when the most luminous galaxy in
    a group is assumed to host the absorption.}
  \label{fig:EWDfits}
\end{figure*}

The group environment sample contains only three groups in which only
an upper limit can be measured on the {\MgII} absorption equivalent
width. We calculated the covering fraction, $f_c$, of the group
environment and isolated galaxy samples for comparison, where we
define the covering fraction as the fraction of absorbers with
$W_r(2796)$ measurements to the total sample ($W_r(2796)$ measurements
and upper limits). Upper limits are considered non-detections
regardless of their value. The uncertainties on $f_c$ are calculated
using the formalism for binomial statistics
\citep[see][]{gehrels86}. The covering fraction of the group
environment sample is $f_c=0.89_{-0.09}^{+0.05}$ compared to
$f_c=0.68_{-0.03}^{+0.03}$ for the isolated sample, a $2.2\sigma$
difference. If we consider the groups J113007$-$144927, $z=0.328$ and
J022815$-$405714, $z=0.2678$ as non-absorbers due to having equivalent
widths smaller than the equivalent width sensitivity limit of
$0.04$~{\AA}, then the covering fraction reduces to
$f_c=0.82_{-0.10}^{+0.07}$, a $1.3\sigma$ difference. Group
environments may be more likely to have associated {\MgII} absorption
than galaxies in isolation, although the result is only marginally
significant. Note that \citet{chen10a} examined a galaxy-selected
sample and found only one non-absorbing group out of eight groups,
which gives a covering fraction of $f_c=0.87_{-0.23}^{+0.10}$ and is
consistent with the values we obtain.

We also tested whether the galaxy properties for the group sample were
any different from the isolated sample. KS tests comparing the
redshifts, $B-$band luminosities, and $B-K$ colors (where available)
of the group sample to the isolated sample show that the two samples
are likely drawn from the same population ($<3\sigma$). Conversely,
the distributions of impact parameters for the group environment
sample result in a significant KS test at the $3.4\sigma$ level,
indicating that the null hypothesis that the two samples are drawn
from the same population is disfavored. The group sample is located at
larger impact parameters on average. However, note that the group
sample in this case includes {\it all} group galaxies. If only one
galaxy in the group actually hosts the absorption, regardless of
whether it is the nearest galaxy or the most luminous, the KS test
indicates that the impact parameter distributions between the group
and isolated samples are likely drawn from the same population.

Since it is difficult to pinpoint which galaxy is giving rise to the
observed absorption, several previous works have either assumed that
the nearest galaxy \citep[e.g.,][]{sdp94, schroetter16}, or the most
luminous/massive galaxy \citep[e.g.,][]{bordoloi11, schroetter16} is
the host galaxy. We further investigate the equivalent width
anti-correlation with impact parameter by assuming that the absorption
is either due to the nearest galaxy to the quasar sightline (projected
distance) or the most luminous galaxy.

\subsection{$W_r(2796)$ vs. $D$: Nearest Galaxy}

Selecting the nearest galaxy to the quasar as the source of the
observed absorption has a historical basis, where \citet{sdp94}
searched for galaxies giving rise to absorption by moving outwards in
$D$ and stopping with the first galaxy at an appropriate
redshift. More recent work has conducted blind (to absorption) surveys
of galaxies with nearby quasar spectra \citep[e.g.,][]{chen10a,
  werk13}. Given the $W_r(2796)$--$D$ anti-correlation and the fact
that the covering fraction decreases with increasing impact parameter
({\magiicat} II), both for isolated galaxies, the nearest galaxy is
more likely to give rise to the absorption, especially since the
{\MgII} CGM radius is $\lesssim200$~kpc.

Figure~\ref{fig:EWDfits}(a) presents the $W_r(2796)$ vs $D$
anti-correlation for isolated galaxies (gray points and arrows) and
group galaxies (square points), where $D$ for the group environments
is selected from the nearest galaxy to the quasar sightline in
projected distance. The nearest galaxy for each group environment is
shown in Figures~\ref{fig:radecspec} and \ref{fig:nospecradec} and the
RA/Dec offsets and impact parameters for each galaxy are listed in
Tables~\ref{tab:obsprops} and \ref{tab:calcprops}, respectively.

To test if there is an anti-correlation between equivalent width and
impact parameter, we ran a non-parametric Kendall $\tau$ rank
correlation test on $W_r(2796)$ against $D$ for all of the square
points in Figure~\ref{fig:EWDfits}(a), accounting for upper limits on
absorption. We found a marginally significant result of $2.9\sigma$,
indicating that the two values may be anti-correlated, and that the
equivalent width of absorption may decrease with increasing impact
parameter. This is in contrast to the highly anti-correlated isolated
sample with $7.9\sigma$ ({\magiicat} II). The CGM of group galaxies
may have a flatter equivalent width profile than isolated
galaxies. However, note that historically, this anti-correlation has
not always been significant in the isolated sample. Only with larger
samples \citep[e.g.,][{\magiicat} II]{chen10a, kcems11} has the
anti-correlation become statistically significant. Also note that the
group environment sample has very few fields where only an upper limit
on absorption can be measured, potentially biasing the sample to a
flatter distribution. A larger group sample would be useful to
investigate the level of bias and better determine how commonly group
environments do not have associated {\MgII} absorption.

To test this further, we parameterized the nearest-galaxy group
environment sample anti-correlation with the Expectation-Maximization
maximum-likelihood method \citep{wolynetz79}, accounting for upper
limits on $W_r(2796)$. We fit a log-linear model similar to that for
the isolated galaxies from {\magiicat} II ($\log W_r(2796) = (-0.015
\pm 0.002)\log D + (0.27 \pm 0.11)$, gray solid and dashed lines). The
group environment fit is shown as the cyan solid line, with $1\sigma$
uncertainties about the fit as dashed lines. The adopted fit to the
group sample is $\log W_r(2796) = (-0.010 \pm 0.003) \log D + (0.35
\pm 0.42)$. This slope is slightly flatter than for the isolated
sample ($1.4\sigma$) but the uncertainties are larger. The fit to the
group data is consistent with the isolated sample within uncertainties
so we cannot definitively state that the equivalent width profile of
nearest-galaxy group environments is flatter than the isolated CGM. A
larger group environment sample size may decrease the uncertainties on
this fit.

\subsection{$W_r(2796)$ vs $D$: Most Luminous Galaxy}

Assuming the most luminous galaxy is giving rise to the detected
absorption is also reasonable. As we found in \citet{cwc-masses,
  magiicat3}, more massive galaxies have a more extended CGM, where
{\MgII} is regularly found out to $0.3R_{\rm vir}$. Using luminosity
as a proxy for mass, more luminous galaxies should host a CGM that
extends out to larger impact parameters, which we found in {\magiicat}
II. The most massive galaxies in a group will likely have the largest
potential wells, allowing for the galaxy to host a more massive
CGM. The covering fraction profiles also show that more luminous
galaxies have a higher covering fraction than less luminous galaxies
at a given impact parameter ({\magiicat} II). For each group, we
select the most luminous galaxy in the $B$-band. These galaxies are
identified as the largest points in Figures~\ref{fig:radecspec},
\ref{fig:nospecradec}, and \ref{fig:EWD}. The luminosities for each
galaxy are also listed in Table~\ref{tab:calcprops}.

Figure~\ref{fig:EWDfits}(b) presents the $W_r(2796)$ vs $D$
anti-correlation for the most luminous group galaxy members. Point and
line types and colors are the same as those in panel (a). The most
luminous galaxies appear to have an even flatter distribution that
what we found for the nearest galaxy sample. A rank correlation test
(accounting for upper limits) on $W_r(2796)$ vs $D$ results in only
$2.6\sigma$, less than for the nearest galaxy sample, although still
marginally significant. We again fit the data with a log-linear model
using the Expectation-Maximization maximum-likelihood method,
accounting for upper limits on $W_r(2796)$. The adopted fit to these
data is $\log W_r(2796) = (-0.007 \pm 0.002) \log D + (0.33 \pm
0.25)$. The slope for the most luminous galaxies is flatter than for
the isolated galaxy sample ($2.8\sigma$), but the full fit is not
significantly different. Assuming the most luminous galaxy in a group
gives rise to the observed absorption, the group {\MgII} CGM may be
more extended than the isolated CGM.

Since we selected the most luminous group galaxies, there may be
biases causing the flatter fit to the data than with the isolated
sample. However, we ran a KS test comparing the luminosities of the
galaxies in this most luminous group galaxy sample to the isolated
sample and found that the two samples were drawn from the same
population ($1.9\sigma$). We also compared the impact parameters of
the two samples and found no significant difference ($2.2\sigma$).

\subsection{$W_r(2796)$ vs $D$: Superposition Model}
\label{sec:superboot}

Using stacked galaxy spectra to probe the CGM of foreground galaxies,
\citet{bordoloi11} found that the possible extension of the group CGM
distribution can be modeled by a superposition of absorption profiles
associated with individual galaxies. This method assumes that the
individual galaxies are not affected by galaxy--galaxy interactions in
the groups, but that the larger equivalent widths are simply due to
the quasar sightline piercing multiple circumgalactic media. To test
this, the authors summed the equivalent widths associated with
isolated galaxies according to the modeled fit to the isolated
galaxies on the $W_r(2796)-D$ plane and the impact parameter
distribution of the group members in question. Because the resulting
superposition model is consistent with the group data, they suggested
that the observed absorption is simply due to a superposition of
individual halos and that the group environment may not affect the
{\MgII} CGM of individual galaxies. We investigate this further using
our distribution of {\magiicat} isolated galaxies.

\begin{figure}[t]
  \includegraphics[width=\linewidth]{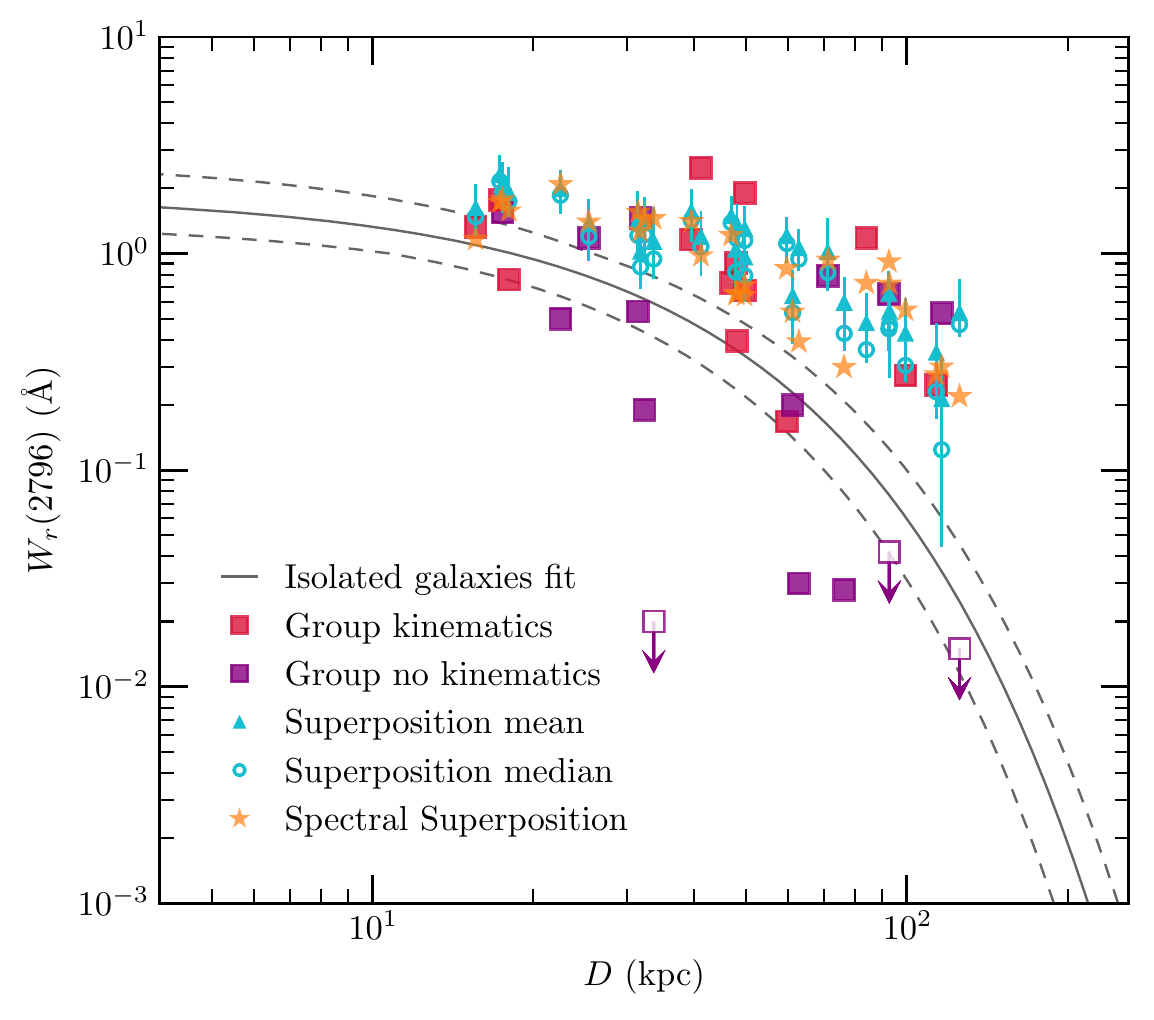}
  \caption[]{Superposition model for each group environment, where the
    impact parameter is defined by the group galaxy nearest to the
    quasar sightline. The choice of galaxy for the impact parameter
    does not affect the results. Gray lines and square points are
    plotted as in Figure~\ref{fig:EWDfits}(a). Cyan symbols and lines
    are the summed equivalent widths from the CGM of multiple galaxies
    (see Section~\ref{sec:superboot} for details). Triangles are the
    mean equivalent width values and the circles the median values for
    1000 realizations in a bootstrap analysis. Vertical blue lines
    indicate the $1\sigma$ uncertainty around the mean equivalent
    width from the bootstrap analysis to take into account the scatter
    in the isolated galaxy sample. Orange stars are superposition
    model equivalent widths calculated from summed absorption profiles
    (see Section~\ref{sec:kinsuperpos}). The superposition model fits
    well with some of the group environments, but misses the lowest
    equivalent width groups.}
  \label{fig:superboot}
\end{figure}

For each group, we substitute equivalent width measurements from
isolated galaxies within similar impact parameters to remove the
potential impact of galaxy--galaxy interactions on the observed
absorption profiles. We first identify galaxies from our isolated
galaxy catalog within $\pm 8$~kpc of each group galaxy member. This
impact parameter range was selected to be as small as possible so that
the $W_r(2796)-D$ anti-correlation does not change drastically over
the $D$ range, but large enough to contain at least five isolated
galaxies. With this sample, we randomly draw an isolated galaxy within
the impact parameter range for each group galaxy member and sum the
associated equivalent widths with the assumption that upper limits on
absorption are ``absorbers'' at the measured upper limit value. This
is done 1000 times for each group using a bootstrap analysis in which
we randomly draw new isolated galaxy replacement equivalent widths for
each realization, and the mean and median of the summed equivalent
widths and $1\sigma$ uncertainties from all of the bootstrap
realizations are calculated. This method therefore takes into account
the spread in the isolated galaxy distribution on the equivalent
width--impact parameter plane, and does not depend on the fit applied
to the isolated sample in this plane \citep[as is the case
  in][]{bordoloi11}.

The results of this superposition model are shown in
Figure~\ref{fig:superboot}, where the point colors and types are
similar to those in Figure~\ref{fig:EWDfits}(a). The choice of plotted
galaxy impact parameter does not affect the results of this analysis
because we are comparing total equivalent widths and take into account
the group galaxy member impact parameters in the equivalent width
summation. Therefore, we choose the nearest galaxy for simplicity. The
cyan triangles (circles) are the mean (median) equivalent width of the
bootstraps for the superposition model, while the vertical lines
indicate the $1\sigma$ uncertainties in the bootstraps to show the
range in possible summed equivalent widths. The superposition model
fits half of the data well, but misses the lower equivalent width
groups. The model points still lie within the scatter of the isolated
points, but tend toward the upper right portion of the
distribution. Given that the model does not explain all of the groups,
especially those with low equivalent widths, it is likely that not all
group member galaxies contribute to the absorption in all cases.

Because summing equivalent widths does not take into account the
reality that gas associated with multiple galaxies may be located at
the same line-of-sight velocities, and therefore the model equivalent
widths may be overestimated, we also calculate superposition model
equivalent widths by summing absorption spectra (for full details, see
Section~\ref{sec:kinsuperpos}). This method accounts for
galaxy--galaxy velocity separations due to slightly different galaxy
redshifts across the group and for absorber--galaxy velocity
separations due to gas motions around individual galaxies. The
resulting summed equivalent widths are presented as orange stars in
Figure~\ref{fig:superboot}. There are some variations in the
calculated values due to the fact that we can only use the subset of
isolated galaxies for which we have the associated quasar
spectrum. Regardless, the equivalent widths derived from the
absorption spectra are similar to those derived by summing equivalent
width values.

With this superposition modeling, we also investigated the expected
covering fraction, $f_c$, of the group environment sample by keeping
track of the number of absorbers and non-absorbers (upper limits on
absorption) in each bootstrap realization. For a group in the
superposition model to be a non-absorber, all galaxies in that group
must not have measurable absorption, i.e., upper limits on absorption
must be randomly drawn for every galaxy in the group. For a galaxy to
be classified as an absorber, at least one galaxy must have
absorption. The mean covering fraction expected from this model is
$f_c=0.83_{-0.01}^{+0.03}$, where the uncertainties are $1\sigma$
uncertainties in the bootstrap realizations from the mean. The value
is comparable to that found for the actual group environment sample
within uncertainties ($f_c=0.89_{-0.09}^{+0.05}$), but is
significantly larger than the isolated galaxy sample
($f_c=0.68_{-0.03}^{+0.03}$). This suggests that absorption is more
likely to be found in group environments in a superposition model than
for isolated galaxies alone. The result that the superposition
covering fraction is lower than the actual value (despite being within
uncertainties) also suggests that the superposition model may be too
simplistic by neglecting galaxy--galaxy interactions.

\begin{figure}[th]
\includegraphics[width=\linewidth]{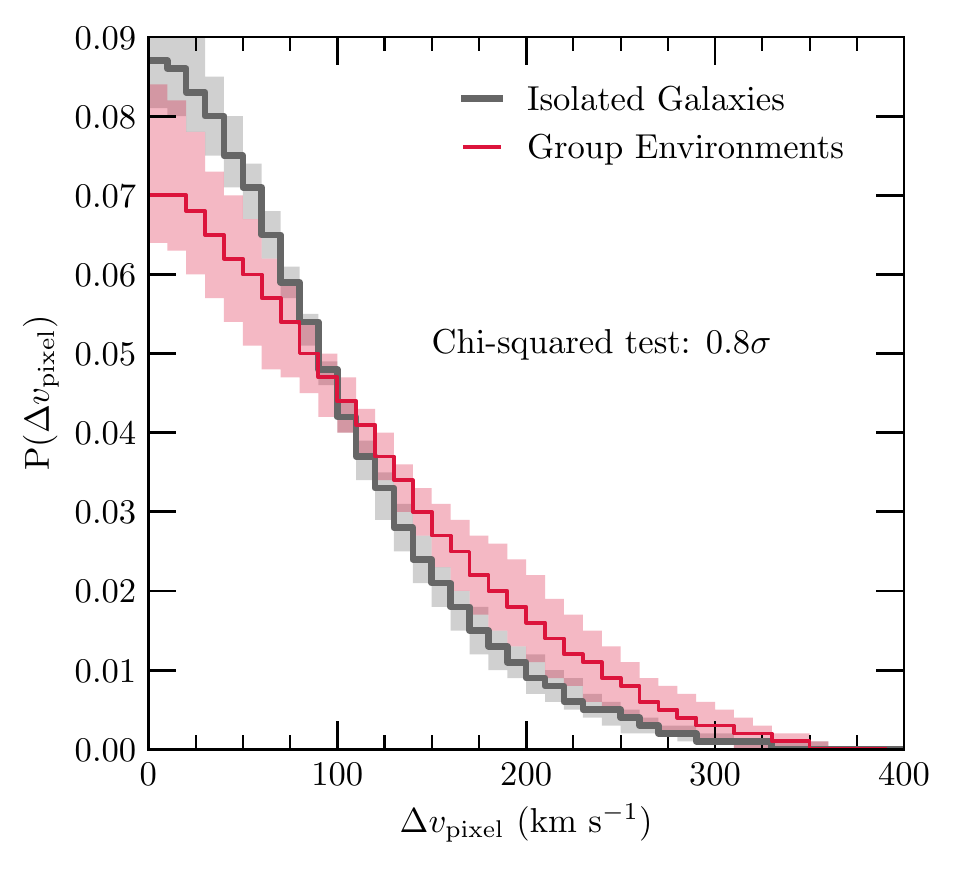}
\caption[]{Pixel-velocity two-point correlation functions (TPCFs)
  comparing absorbers associated with isolated galaxies (thick gray
  line and shading) to absorbers associated with galaxies in group
  environments (thin red line and shading). The samples contain all
  absorber--galaxy pairs in our sample with high-resolution quasar
  spectra. The histograms represent the TPCF, while the shaded regions
  are the uncertainties on the TPCF from a bootstrap
  analysis. Absorbers associated with galaxy groups have statistically
  similar velocity dispersions as those associated with isolated
  galaxies ($0.8\sigma$). However, the group sample has more power at
  $\Delta v_{\rm pixel}\sim 200$~{\kms}, which may be due to the
  superposition of the CGM of multiple galaxies, tidal interactions,
  or intergalactic transfer.}
\label{fig:tpcf}
\end{figure}

\section{Kinematics}
\label{sec:kinematics}

The equivalent width of an absorber is proportional to the number of
clouds fit with Voigt profile modeling \citep[e.g.,][]{pb90, cvc03,
  evans-thesis}. The group galaxies appear to have a more extended
CGM, where group galaxies may have a larger $W_r(2796)$ at a given $D$
than isolated galaxies, at least for scenarios in which the most
luminous group galaxy hosts the observed absorption. This indicates
that the absorber velocity spread, column density (and thus the
metallicity, path length, ionization conditions, etc.), or some
combination may be larger for group environments. Therefore, we
investigate the kinematics of the group absorbers using the
pixel-velocity two-point correlation function (TPCF).

The TPCF is defined as the probability distribution function of the
velocity separation of every absorbing pixel pair in a sample. Full
details of the pixel-velocity TPCF method are published in {\magiicat}
IV \citep[also see][]{magiicat5, nielsen17}. To create the TPCF, we
obtain the pixel velocities in every absorber (defined by the velocity
bounds of absorption, see Section~\ref{sec:qso}) for a sample.
Absorption regions (and their associated pixel velocities) that have
equivalent widths less than our completeness cut of $0.04$~{\AA} are
not included in this analysis. We then calculate the velocity
separations of each possible pixel pair in a given sample, without
repeats.  The absolute value of the velocity separations is
calculated, and these values are binned in 10~{\kms} bins. The count
in each bin is then normalized by the total number of pixel-velocity
pairs in the sample to create a probability distribution
function.\footnote{For the samples presented here, there are roughly 3
  million (isolated galaxy sample) and 500,000 (group environment
  sample) pixel-velocity pairs in the TPCF calculations.} The TPCF is
roughly a measure of the velocity dispersion of absorbers in a given
sample. Note that TPCFs can be created for only those galaxies/groups
in which absorption is detected; nonabsorbers do not provide kinematic
information due to the lack of pixels contributing to observed
absorption.

Uncertainties on the TPCF are calculated using a bootstrap
analysis. We randomly draw, with replacement, the same number of
absorbers from the sample in question and calculate the TPCF for that
realization. We do this for 100 realizations and calculate the mean
and standard deviation of the realizations. The uncertainties we plot
are $1\sigma$ bootstrap uncertainties.

We calculated the TPCF for both our group sample with high-resolution
quasar spectra (red points in Figure~\ref{fig:radecspec}) and for our
isolated galaxy sample with high-resolution quasar spectra presented
in {\magiicat} IV. There are 14 group environments and 46 isolated
galaxies involved in the TPCF calculations. The median redshifts for
the samples are tabulated in Table~\ref{tab:v50}. The TPCFs are
presented in Figure~\ref{fig:tpcf}, where the red curve and shaded
region are the TPCF and uncertainties, respectively, for the group
sample. Isolated galaxies are plotted as a gray curve and shaded
region.


\addtocounter{table}{2}

\begin{deluxetable}{lrlll}
\tablecolumns{5} 
\tablewidth{0pt} 
\tablecaption{TPCF {\vfifty} and {\vninety}
  Measurements \label{tab:v50}}
\tablehead{
  \colhead{Sample} &
  \colhead{\# Gals} &
  \colhead{$\langle z_{\rm abs} \rangle$} &
  \colhead{{\vfifty}\tablenotemark{b}} &
  \colhead{{\vninety}\tablenotemark{b}} 
}
\startdata

Isolated Galaxies  & 46 & $0.656$ & $ 66_{- 4}^{+ 3}$ & $177_{-  9}^{+  9}$ \\[3pt]
Group Environments & 14 & $0.411$\tablenotemark{a} & $ 79_{-11}^{+13}$ & $199_{- 27}^{+ 22}$ \\[3pt]

Galaxy--Galaxy Groups & 10 & $0.443$ & $ 85_{-15}^{+12}$ & $208_{- 35}^{+ 20}$ \\[3pt]
Galaxy--Dwarf Groups  &  4 & $0.330$ & $ 60_{-34}^{+ 8}$ & $139_{- 68}^{+ 20}$ \\[-5pt]

\enddata
\tablenotetext{a}{Median redshift measured only from the group
  galaxies with high-resolution quasar spectra (red points in
  Figures~\ref{fig:radecspec} and \ref{fig:EWD}).}
\tablenotetext{b}{{\kms}}
\end{deluxetable}


From Figure~\ref{fig:tpcf}, we find that absorbers associated with
galaxies in group environments have statistically similar velocity
dispersions as those associated with isolated galaxies, where a
chi-squared test comparing the TPCF distributions of the group galaxy
sample to the isolated galaxies results in a significance of
$0.8\sigma$. We further characterize the TPCFs by determining the TPCF
velocity separation, $\Delta v_{\rm pixel}$, within which $50\%$ and
$90\%$ of the velocity separations are located, {\vfifty} and
{\vninety}, respectively. These values are $ 79_{-11}^{+13}$~{\kms}
and $199_{- 27}^{+ 22}$~{\kms} for the group environment sample,
respectively, and $66_{-4}^{+3}$~{\kms} and $177_{-9}^{+9}$~{\kms} for
the isolated galaxy sample, respectively. These values are also
tabulated in Table~\ref{tab:v50}. The {\vfifty} and {\vninety} values
for the group sample are both larger than for the isolated sample,
although the uncertainties overlap. The TPCFs for both samples
generally extend out to the same velocity separation of
$\sim350$~{\kms}, but the group TPCF has more power at
$\sim200$~{\kms} than the isolated TPCF. Larger velocities would be
expected in a group superposition of halos and/or where interactions
between group galaxies are occurring. We investigate this further in
the following sections.

\subsection{Galaxy--Galaxy Luminosity Ratios}
\label{sec:kinmerger}

If we assume that the absorption properties are due to galaxy--galaxy
interactions, there may be some observable differences due to the type
of environment, which we quantify by calculating the luminosity ratio
of the two brightest galaxies in a group. A majority of the sample
presented here involves pairs of galaxies with similar luminosities
that are close in projection. In these environments, the CGM (and the
galaxies themselves) are expected to be impacted more dramatically by
interactions than environments where there is a large galaxy and one
or more ``dwarf'' galaxies. Thus we investigate this effect by
slicing the sample by the luminosity ratio between the two brightest
galaxies in each group, assuming the $B-$band luminosity is a proxy
for galaxy mass. We define galaxy--galaxy groups as those where the
ratio between the two brightest galaxies (most luminous over
second-most luminous) is $L_1/L_2<3.5$, regardless of the impact
parameter between the two galaxies. Galaxy--galaxy groups may result
in a future major merger. Galaxy--dwarf groups are defined as
group environments where the ratio is $L_1/L_2\geq3.5$, and these may
result in a future minor merger.

\begin{figure}[t]
  \includegraphics[width=\linewidth]{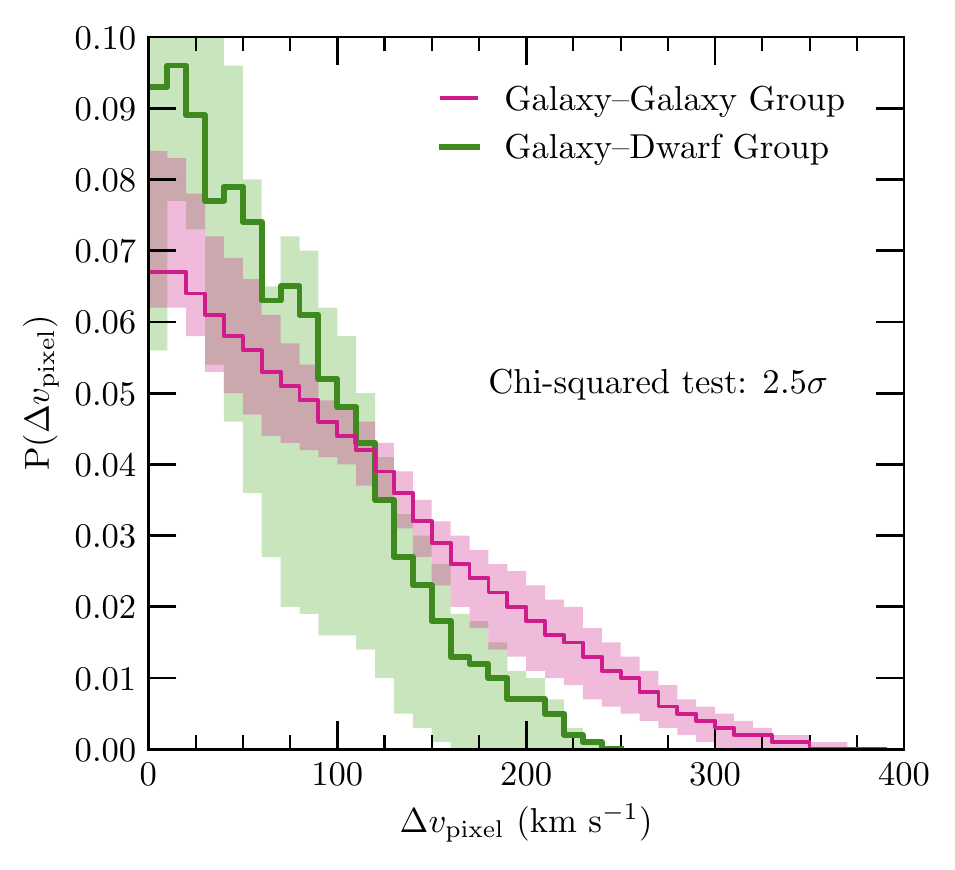}
  \caption[]{Gas kinematics comparing galaxy--galaxy groups ($L_{B,
      {\rm ratio}}<3.5$; pink) and galaxy--dwarf groups ($L_{B, {\rm
        ratio}}\geq3.5$; green). Absorption associated with
    galaxy--galaxy groups may have larger velocity dispersions than
    absorption associated with galaxy--dwarf groups. Although this
    result is only marginally significant at the $2.5\sigma$ level,
    largely due to the galaxy--dwarf subsample containing only four
    groups, the galaxy--dwarf uncertainties trend towards narrower
    velocity dispersions.}
  \label{fig:mergers}
\end{figure}

\begin{figure*}[ht]
  \centering
  \includegraphics[width=\linewidth]{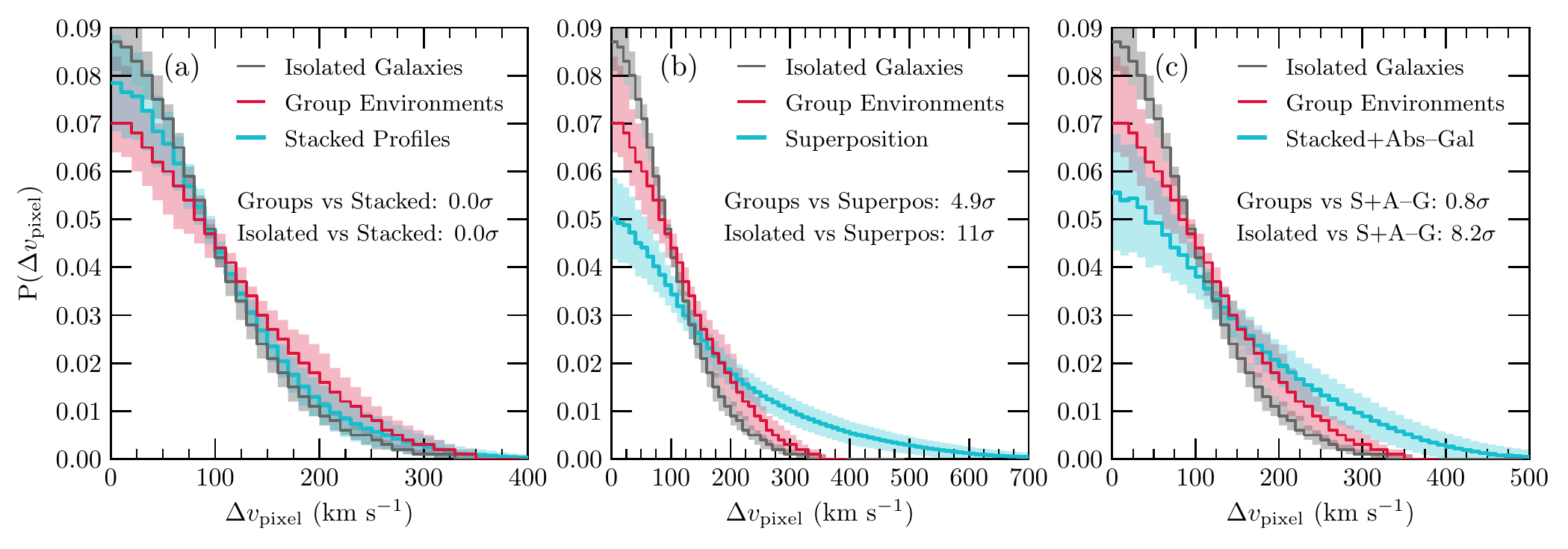}
  \caption[]{Superposition modeling for the group environment
    pixel-velocity TPCF. In each panel, the gray lines and shading
    represent the isolated sample while the red lines and shading
    represent the group sample, both previously plotted in
    Figure~\ref{fig:tpcf}. The cyan lines and shading are the TPCFs
    for three different superposition model cases: (a) Case 1: A
    simple stack of absorption profiles creates a TPCF that is
    consistent with the isolated TPCF ($0\sigma$) and the group TPCF
    ($0\sigma$). (b) Case 2: A proper superposition of halos that
    includes both absorber--galaxy and galaxy--galaxy velocity offsets
    for the model absorption profiles results in a TPCF that is
    inconsistent with both the group TPCF ($4.9\sigma$) and the
    isolated TPCF ($11\sigma$). (c) Case 3: A stack of absorption
    profiles, where each contributing absorber is shifted from a
    common redshift (absorber--galaxy velocity offsets), is comparable
    to the group environment TPCF ($0.8\sigma$) but not the isolated
    galaxy TPCF ($8.2\sigma$).}
  \label{fig:supertpcf}
\end{figure*}

The median equivalent width for galaxy--galaxy groups, $\langle
W_r(2796) \rangle = 0.74\pm0.17$~{\AA} (mean $0.87\pm0.14$~{\AA}), is
$1.7\sigma$ ($1.8\sigma$) larger than for galaxy--dwarf groups,
$\langle W_r(2796)\rangle=0.27\pm0.21$~{\AA} (mean
$0.48\pm0.17$~{\AA}). Out of the three non-absorbing groups in the
sample, two are classified as galaxy--dwarf groups, while one is a
galaxy--galaxy group, resulting in covering fractions of
$f_c=0.95_{-0.11}^{+0.04}$ (galaxy--galaxy) and
$f_c=0.78_{-0.22}^{+0.14}$ (galaxy--dwarf), which are consistent
within uncertainties. These results suggest that the kinematics and/or
column densities of absorbers depends on the group galaxy luminosity
ratio, potentially hinting at interaction/merger effects.

Figure~\ref{fig:mergers} presents the TPCFs comparing galaxy--galaxy
and galaxy--dwarf groups for only the absorbers in the subsamples
(recall that there is no kinematic information in
nonabsorbers). Galaxy--galaxy groups host absorbers with a larger
velocity dispersion than galaxy--dwarf groups, but the result is only
marginally significant ($2.5\sigma$) due to the large uncertainties in
the galaxy--dwarf sample. Regardless, groups in which a major merger
may occur in the future (galaxy--galaxy) appear to drive the kinematic
trends in the group environment TPCF.

\subsection{Kinematics Superposition Modeling}
\label{sec:kinsuper}

If the superposition model presented in Section~\ref{sec:superboot}
and in \citet{bordoloi11} for the equivalent width of absorption
associated with group galaxies is accurate, then the model should also
apply to the kinematics of these absorbers. Here we apply the
superposition technique to create model absorbers and use these to
calculate TPCFs for three different cases. These cases include (1) the
absorption is ``stacked'' where the absorption due to multiple
galaxies all lies at the same redshift ($z_{\rm abs}$); (2) the
absorption is truly associated with individual galaxies, where the
kinematics depend on both the galaxy--galaxy velocity distributions of
each group and the absorber--galaxy velocity distribution expected for
each member galaxy to reflect their individual baryon cycles; and (3)
the absorption is due to a common intragroup medium in which the gas
is observed at a common velocity with small velocity offsets due to
random gas motions.

In each case we work only with those absorbers (and upper limits on
absorption) for which we have high-resolution quasar spectra in order
to obtain the detailed kinematics. For instances where only an upper
limit is measured in these spectra, the (non-)absorption does not
contribute to the TPCF because there are no pixels contributing to
absorption in these cases. Though note that the covering fraction of
the individual group member contributions is often less than one, and
in many cases no absorption is modeled for entire group environments.

\subsubsection{Case 1: Stacked Profiles TPCF}

For the first case, we created stacked absorption profiles by randomly
selecting isolated galaxies within $\pm 8$~kpc of each galaxy in a
group and obtained the absorption profiles associated with each. We
rebinned every profile onto a common velocity array with 3~{\kms}
pixel widths to match the resolutions of the HIRES and UVES
spectrographs. These rebinned absorption profiles are then summed in
velocity space, where $v=0$~{\kms} is the optical depth-weighted
median of the summed absorption ($z_{\rm abs}$). This assumes that the
individual group galaxy absorption contributions are centered at the
same redshift regardless of the spread in group galaxy redshifts or
any offset the absorption might have from the host galaxy. This may be
interpreted as an intragroup medium with more absorbing material at a
given line-of-sight velocity than in an isolated environment. Each
group has its own summed absorption profile with contributions from
each group member galaxy. The summed absorption profiles for each
group were then used to calculate a TPCF. This analysis was done for
1000 bootstrap realizations where the random selection of isolated
galaxies that go into the superposition model are bootstrapped, and
the mean and standard deviation in each bin of the TPCF realizations
were calculated.

Figure~\ref{fig:supertpcf}(a) presents the isolated and group TPCFs
from Figure~\ref{fig:tpcf}, with the addition of the stacked profiles
TPCF in cyan. The mean of the stacked profiles TPCF bootstrap
realizations is plotted as the cyan line, while the $1\sigma$ standard
deviation of the realizations is plotted as the cyan shaded region. We
find that this ``stacked'' TPCF is consistent with both the isolated
sample (chi-squared test: $0\sigma$) and the group sample ($0\sigma$),
though it is still narrower than the group sample. The larger
uncertainties on the stacked TPCF compared to the isolated TPCF,
despite being drawn from the same samples, is likely due to the random
nature of the analysis and the smaller group galaxy sample size
compared to the isolated sample (14 versus 46). This stacked profiles
model is a useful exercise since it represents the minimum velocity
spreads possible in a superposition scenario. However, this model is
unrealistic because it neglects the relative motions of gas around
individual galaxies due to baryon cycle processes, as well as the
relative velocities between group member galaxies. Therefore, the
model is ruled out.

\subsubsection{Case 2: Superposition TPCF}
\label{sec:kinsuperpos}

In the second case, we conduct a similar analysis as the previous
section, but now adopt realistic galaxy and gas velocity
shifts. Before we sum the individual absorbers, the absorbers are
shifted in velocity for both: (1) Absorber--galaxy velocity offset
based on the Gaussian distribution of velocity offsets presented in
\citet{chen10a}, with $\langle v_{\rm abs-gal} \rangle =16$~{\kms} and
$\sigma_{\rm abs-gal}=137$~{\kms}; and (2) Galaxy--galaxy velocity
offset based on the distribution of group galaxy redshifts. The
redshift of the group galaxy with the smallest impact parameter in the
field defines $v=0$~{\kms} for simplicity, with additional galaxies
having velocity offsets from that. We randomly draw absorber--galaxy
velocity offsets from the \citet{chen10a} Gaussian distribution for
each group galaxy. These velocity shifts combined more accurately
represent the distribution of gas expected if the absorption is truly
associated with individual galaxies in the group and if the gas is not
influenced by or coupled to other group members. The group member
absorption profile contributions are then summed, and the total
absorption redshift, $z_{\rm abs}$, and absorption velocity bounds are
recalculated. The TPCF analysis then proceeds as above.

The result of this analysis is presented in
Figure~\ref{fig:supertpcf}(b). The superposition TPCF is plotted in
cyan, while the isolated and group samples are plotted as before. The
resulting TPCF has a velocity dispersion that is much too large
compared to the true group sample (chi-squared test
result:~$4.9\sigma$). If we do not shift the absorbers according to
the absorber--galaxy velocity offset distribution (velocity shift
number 1 above), the TPCF comparison is slightly more extended and
inconsistent with the group TPCF at the $5.0\sigma$ level. This
exercise suggests that the hypothesis in which each group galaxy may
contribute separately to the observed absorption profile is
incorrect. This is largely due to the spread in group galaxy redshifts
and indicates that the observed gas is coupled to the group
environment or 1-2 galaxies rather than every individual galaxy in the
group. Thus the superposition model appears to be incorrect.

\subsubsection{Case 3: Absorber--Galaxy Velocity Offsets TPCF}

The third case assumes that there is a common intragroup medium in
which multiple galaxies contribute gas but the individual
contributions are offset slightly from a common redshift. In this
case, we assume all contributing absorbers start with $v=0$~{\kms}
representing $z_{\rm abs}$ for each absorber (like the stacked spectra
TPCF above), and then randomly shift these velocities individually
according to the absorber--galaxy velocity offset presented by
\citet{chen10a}. Then we sum the absorption profiles, redefine $z_{\rm
  abs}$ and velocity bounds for the new summed profile, and then
proceed with the TPCF calculation. This method is similar to the
superposition model in the preceding section, except we do not include
galaxy--galaxy velocity separations, which dominate the kinematic
spread.

The resulting TPCF is presented in Figure~\ref{fig:supertpcf}(c),
where lines, shading, and colors are similar to the previous
panels. The stacked, absorber--galaxy velocity offset TPCF (presented
as the cyan line and shading) is comparable to the group environment
sample, with a chi-squared test result of $0.8\sigma$, however the
tail on the model TPCF appears to be too extended. Compared to the
isolated sample, the model TPCF has a velocity dispersion that is too
large ($8.2\sigma$). This result and the previous two superposition
model scenarios suggest that an intragroup medium as the physical
region giving rise to the observed {\MgII} gas is more plausible than
a true superposition of galaxy halos model.

\subsubsection{Other TPCFs}

\begin{figure}[t]
  \includegraphics[width=\linewidth]{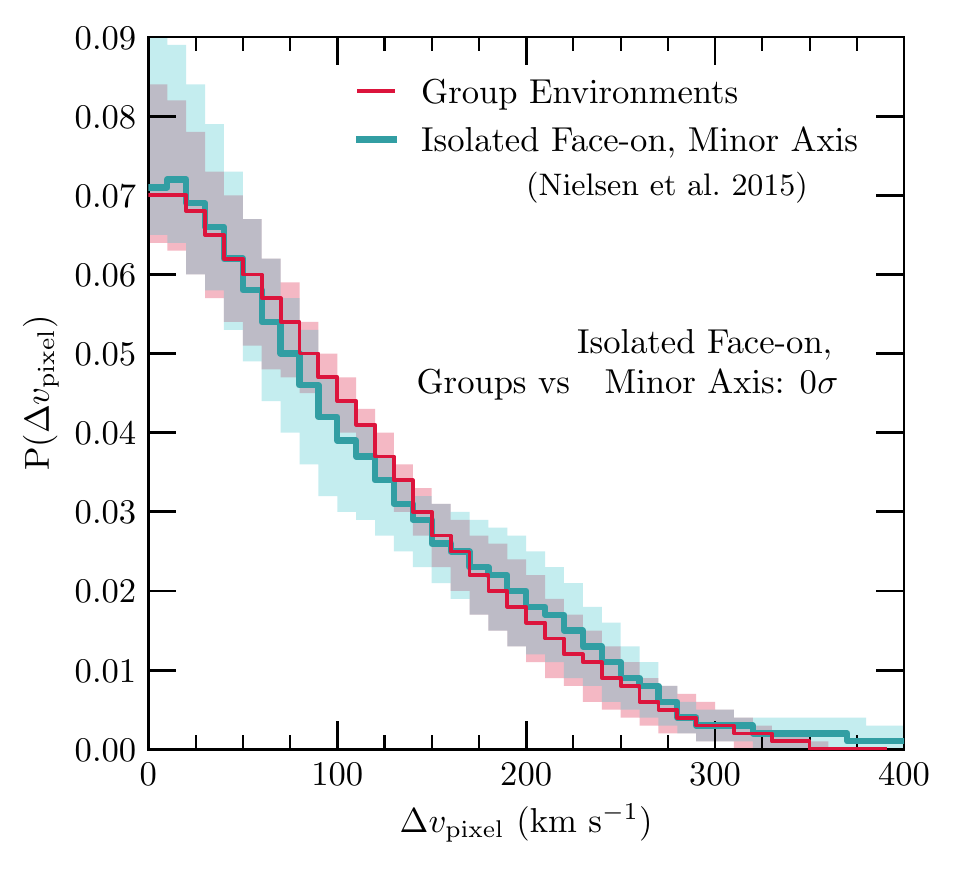}
  \caption[]{TPCF for absorbers associated with isolated face-on
    ($i<57^{\circ}$), minor axis ($\Phi\geq45^{\circ}$) galaxies from
    {\magiicat} V (cyan thick line and shading) compared to the full
    group environment (red line and shading) sample. The face-on,
    minor axis galaxy sample is a subset of the full isolated galaxy
    sample where the subsample's kinematics were associated with
    outflows in {\magiicat} V. The kinematics for absorbers found in
    group environments are comparable to those along the minor axis of
    face-on, isolated galaxies. This suggests that the gas probed by
    {\MgII} in group environments is either outflowing material, or is
    agitated similarly to outflows (potentially streams from tidal
    stripping).}
  \label{fig:supertpcfminor}
\end{figure}

Curiously, the group environment TPCF is consistent with presumably
outflowing gas in the isolated galaxy subsamples published in
\citet[][hereafter {\magiicat} V]{magiicat5}. These subsamples are
subsets of the isolated galaxy catalog presented as gray symbols and
lines in all figures so they have galaxy properties (redshifts, impact
parameters, luminosities, etc.) that are consistent with the group
environment sample. In Figure~\ref{fig:supertpcfminor} we present a
comparison between the full group environment sample (red) and the
isolated face-on, minor axis sample (cyan) from {\magiicat} V. The
latter subsample consists of face-on galaxies ($i<57^{\circ}$) probed
along the minor axis ($\Phi\geq45^{\circ}$) by the quasar sightline,
which is expected to be the orientation at which outflows are best
measured. A chi-squared test comparing this TPCF and the group
environment sample results in a significance level of $0.0\sigma$. Due
to the complexity of gas flows in group and interacting environments,
it is unlikely that the gas observed in the group environment sample
is (solely) due to outflowing gas, especially in a statistical sense
as is the case for the TPCFs. However, this TPCF comparison does
suggest that the processes responsible for the properties of this
group gas may disturb the gas similarly to outflows through tidal
stripping, or even induce outflows.

In {\magiicat} V, we explored several more subsamples sliced by galaxy
orientation properties and galaxy color to better understand the
processes traced by {\MgII} absorption. For orientations in which
outflows are expected to dominate the observed absorption signatures,
the kinematics are consistent with the group sample. For those
orientations in which outflows are non-existent or where accretion is
expected to dominate, the kinematics are inconsistent with the group
environment sample.

\section{Cloud Column Densities and Velocities}
\label{sec:NvsV}

\begin{figure}[t]
  \includegraphics[width=\linewidth]{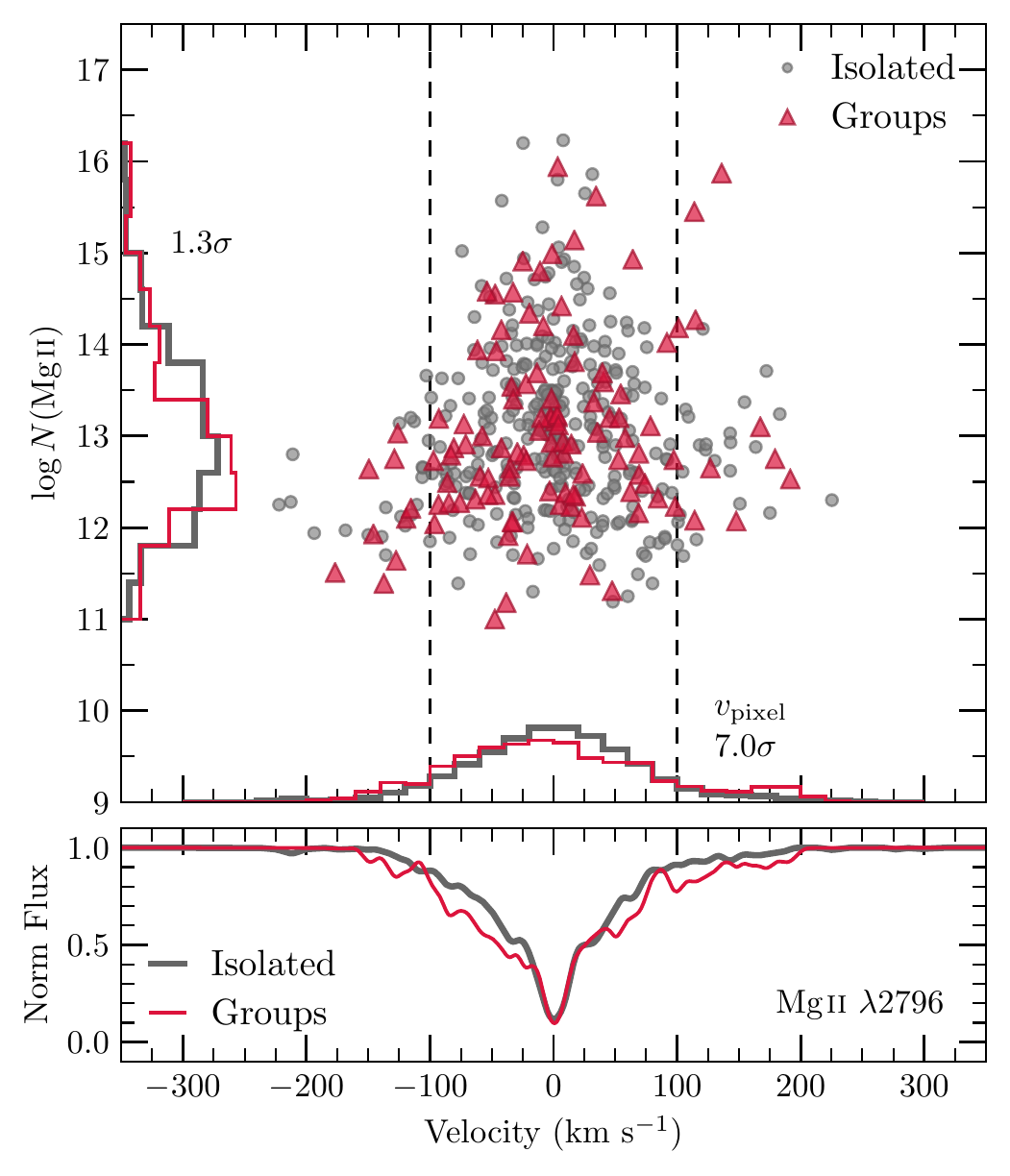}
  \caption[]{(Top) VP model cloud column densities and velocities
    comparing group environments (red triangles and thin lines) to
    isolated galaxies (gray circles and thick lines). Histograms
    compare the distributions of the cloud column densities (left
    axis) and pixel velocities (bottom axis) for the two samples,
    normalized by the number of points in each sample. Vertical dashed
    lines at $v=\pm 100$~{\kms} are plotted to guide the eye.  The
    cloud column densities are comparable between the group and
    isolated samples, but the group sample has a significantly higher
    fraction of $v\geq100$~{\kms} clouds ($16.8\%$) than the isolated
    sample ($13.5\%$). (Bottom) Average model spectra for the group
    environments (red) and isolated galaxies (gray).}
  \label{fig:NvsV}
\end{figure}

To examine the ``clumpiness'' of the absorbers along the line of
sight, we plot the column densities and velocities of each VP fitted
cloud component in the top panel of Figure~\ref{fig:NvsV}. Red
triangles represent the VP modeled clouds for the group sample with
high-resolution quasar spectra and gray circles are those for the full
isolated sample from {\magiicat} IV. The left histograms show the
distribution of cloud column densities for the two samples, while the
bottom histograms show the distribution of pixel velocities (note that
the points in the scatter plot show {\it cloud} velocities, which are
represented by the red ticks at the top of the absorption profile
panels in Figure~\ref{fig:radecspec}). Showing the pixel velocities
gives a more accurate picture of the velocity spread of the absorbers
and are the values used to calculate the TPCFs. In both histogram
sets, thin red lines represent the group sample and thick gray lines
are the isolated sample.

Overall, the VP model cloud column densities and velocities for the
group sample do not differ significantly from the isolated sample. The
highest velocity clouds tend to have small column densities and the
highest column density clouds have the smallest velocities, a result
that largely reflects the velocity zero point definition (absorption
redshift). There is the exception of a few group sample clouds at
$v_{\rm pixel}\geq100$~{\kms} and $\log N({\MgII})=14-16$. A KS test
comparing the cloud column density distributions indicates that the
two samples are drawn from the sample population at the $1.3\sigma$
level. Lower limits on the column densities are considered
measurements at the value of the limit.

However, the pixel velocities for the two samples are different: an
$F$-test comparing the variance in the distributions rules out the
null hypothesis that the two samples are drawn from the same
population at the $7.0\sigma$ level. The pixel velocities for the
group sample have a flatter distribution and are more extended than
for the isolated sample, similar to the TPCFs.

The group sample also has a significantly ($3.3\sigma$) higher
fraction of ``high velocity'' ($v \geq 100$~{\kms}) clouds than the
isolated sample, with $16.8_{-0.8}^{+1.7}\%$ for the groups compared
to $13.5_{-0.9}^{+0.6}\%$ for the isolated sample. The $1\sigma$
uncertainties on these fractions were calculated by conducting a
bootstrap analysis over 1000 realizations in which cloud velocities
from each sample were randomly drawn with replacement and new
fractions were determined. For only galaxy--galaxy groups
($L_1/L_2<3.5$), the fraction increases to
$19.5_{-1.1}^{+2.3}\%$. These ``high velocity'' clouds contribute to
the increased number of pixel velocity separations of
$\sim100-200$~{\kms} in the group TPCF compared to the isolated
TPCF. However, the group and isolated samples have similar numbers of
clouds per absorber on average, with $n_{\rm clouds} = 8.1\pm1.1$ for
the groups sample compared to $n_{\rm clouds} = 7.1\pm0.7$ for the
isolated sample. Restricting the group sample to galaxy--galaxy
groups, we find $n_{\rm clouds} = 8.7\pm1.4$, which is larger but
still consistent within uncertainties.

To test if there is more material along the line of sight in group
environments (i.e., the absorbing gas has a larger physical
distribution, has a higher density, or some combination of the two)
due to probing the CGM of two or more galaxies, we compare the total
column densities of absorbers found in group environments to those in
isolated environments. The median (mean) total column densities are
$\log N({\MgII})=14.20\pm0.32$ ($14.25\pm0.26$) for groups and $\log
N({\MgII})=13.89\pm0.18$ ($14.21\pm0.15$) for isolated galaxies. These
values are consistent within uncertainties. For only galaxy--galaxy
groups, we find $\log N({\MgII})=14.40\pm0.41$ ($14.41\pm0.33$), which
is still consistent within uncertainties with the isolated sample. If
the quasar line-of-sight is probing two or more galaxies as expected
in a superposition model, we would expect the group environment column
densities to be about 0.3~dex larger than the isolated sample
(assuming the absorption from both halos have similar column
densities). This may not be the case (though note that the
uncertainties are also $\sim0.3$~dex), which could either indicate the
individual halos contribute different amounts of gas or the
superposition model is incorrect. A KS test comparing the total column
density distributions for the group (galaxy--galaxy group) and
isolated samples results in a significance of $0.03\sigma$
($0.2\sigma$), thus we cannot rule out that the two samples were drawn
from the same population. This indicates that the amount of material
observed along the line of sight may be similar in group and isolated
environments.

Finally, we plot the average model absorption spectra for the group
(thin red line) and isolated (thick gray line) samples in the bottom
panel of Figure~\ref{fig:NvsV}. We use the model spectra (red lines in
Figure~\ref{fig:radecspec}) to remove any contributions to the average
spectra from noise and blends. Comparing the two samples, we find that
the average absorption spectra are similarly concentrated at
$v=0$~{\kms}, but the average group absorption spectrum has more
optical depth on average at higher velocities, particularly
$v\gtrsim100$~{\kms}. The reader may be concerned that higher velocity
components in the isolated sample are washed out due to averaging the
absorption spectra for 46 absorbers, whereas the group sample is only
averaging 14 absorbers. However, a bootstrap analysis on the isolated
sample with 5,000 realizations of 14 randomly drawn isolated absorbers
(without replacement) found that these realizations are rarely
consistent with the group average absorption spectrum. In fact,
$3\sigma$ bootstrap uncertainties on the average absorption profile
for isolated galaxies are plotted, but they are on the order of or
smaller than the line thickness. Therefore, the dilution of isolated
galaxy absorption features does not appear to be an issue. However, a
larger group environment sample would be beneficial to further examine
this.

\section{Discussion}
\label{sec:discussion}

The previous sections show that, statistically, {\MgII} absorbers in
group environments have absorption properties that are largely
comparable to their isolated counterparts within uncertainties. The
median equivalent width is $1.7\sigma$ larger than for isolated
galaxies and the anticorrelation between equivalent width and impact
parameter may be flatter depending on which galaxy is assumed to host
the absorption. Group environments have larger CGM covering fractions
than isolated galaxies ($2.2\sigma$). The kinematics of gas in group
environments have similar velocity dispersions compared to those in
isolated environments, although the group sample has a higher fraction
of high-velocity clouds (VP components) fitted to the absorbers. Group
absorbers have more optical depth at larger line-of-sight
velocities. Finally, the velocity dispersions and median equivalent
widths for galaxy--galaxy groups ($L_1/L_2<3.5$) are marginally larger
than for galaxy--dwarf groups ($L_1/L_2\geq3.5$), although the
covering fractions are consistent.

To better understand the underlying physics involved, we tested the
superposition model of \citet{bordoloi11} on equivalent widths and
kinematics and found that this model generally appears to explain the
larger equivalent width systems in the group sample. When studying the
absorber kinematics in a superposition model, simply stacking
absorption profiles appears to model the group TPCF extended velocity
dispersion. However, the resulting TPCF is unrealistic since it
neglects both the relative velocities between group member galaxies
and the absorber--galaxy velocities due to baryon cycles associated
with individual galaxies. Therefore, we rule this model out. A proper
kinematic superposition of CGM gas in which these velocity shifts are
accounted for results in velocity dispersions that are much too
large. These two models bracket the group sample and indicate that the
superposition model is too simplistic, especially since group
environments likely have the added complication/confusion of
galaxy--galaxy interactions.

Previous work looking at individual group environments favored various
scenarios giving rise to the observed absorption. For example,
\citet{ggk1127} found two groups in the Q1127$-$145 field. For the
larger equivalent width group at $z_{\rm abs}=0.313$, the authors
suggested that the absorption was due to tidal tails and streams
bridging the group galaxies. This is supported by the observation of
perturbed morphologies for three of the brightest galaxies in the
group, with possible tidal streams extending out to at least
$\sim25$~kpc in deep {\it HST} imaging. For the other group in the
field at $z_{\rm abs}=0.328$, the galaxies do not appear to have
perturbed morphologies, have similar metallicities, and the gas has a
low {\MgII} equivalent width. The origin of this weaker absorption is
therefore ambiguous and the authors did not assign any scenario to
explain this gas.

\citet{whiting06} found eight galaxies associated with strong
absorption, all of which appeared to be early-type galaxies. The
authors concluded that absorption associated with so many early-type
galaxies was rare, and they could not rule out intragroup gas as the
source of absorption. In their preferred scenario, galaxy interactions
remove gas from the individual galaxies and deposit the gas into an
intragroup medium. More recently, \citet{bielby17} identified five
galaxies in MUSE observations associated with a strong absorber. The
authors also preferred an intragroup medium scenario in which the gas
is accreting onto the overall group halo, and suggested that this
material may have been sourced from the accretion/outflow of material
from individual galaxies that mixed into the group environment. This
latter scenario is described as a ``superposition'' by the authors,
but one in which galaxy interactions do not contribute to the overall
intragroup halo.

By studying the environments of two previously known isolated
absorbers with MUSE, both \citet{peroux17} and \citet{rahmani18} found
additional galaxies at the redshift of the known absorbers. In the
former work for the Q2128$-$123 field, one galaxy in the group is
significantly more luminous than the rest ($L_1/L_2=56$, a
galaxy--dwarf group here) and at the lowest impact parameter to the
quasar sightline. The authors found that the gas was largely
associated with this most luminous, nearest galaxy, either as
co-rotating halo material, and/or as accretion. They also suggest that
some portion of the observed gas is associated with an intragroup
medium. In the latter work, \citet{rahmani18}, the authors studied the
Q0150$-$202 field and also conclude that the observed absorption is
associated with the galaxy nearest to the quasar sightline, although
it is not the most luminous in the group. Based on the gas kinematics
and galaxy morphology information, the authors conclude that this gas
is also co-rotating and potentially accreting in a warped disk. Both of
these absorbers have {\MgII} equivalent widths significantly less than
the median equivalent width of the group sample, where the
superposition model in Figure~\ref{fig:superboot} does not match the
observed equivalent widths.

Based on the results presented in the previous sections and
considering that the absorber--group pairs detailed in the previous
paragraphs are included in the present sample, we also support an
intragroup medium scenario where one or more galaxies contribute
material, but also one in which galaxy interactions play some part in
distributing the gas throughout the group halo rather than a general
superposition of multiple galaxy halos scenario. The degree to which
each of these contributions participate in shaping the intragroup
material largely depends on individual circumstances of the groups in
question as shown above. However, we are examining the impact of the
group environment in a statistical manner and are less concerned with
the particulars, which we leave to other work. We arrive at our
favored scenario for the following reasons:

First, the $W_r(2796)-D$ superposition model in
Section~\ref{sec:superboot} generally agrees with the equivalent
widths for the largest equivalent width groups. This would indicate
that the largest equivalent width absorbers have (on average) larger
column densities, larger velocity spreads, or some combination of both
due to probing multiple unrelated halos of gas. However, the median
total column densities for the isolated and group samples for which we
have quasar spectra (basically the kinematics subsample) are
consistent within uncertainties. Additionally, the kinematics in
Section~\ref{sec:kinematics} show that the group environment and
isolated galaxy TPCFs are consistent within uncertainties, although
there is increased optical depth at larger velocities in the group
sample. Examining the group environment sample in more detail, the
kinematics may depend on the luminosity ratio of the two brightest
galaxies in the group (the result is only marginally significant due
to large uncertainties in the galaxy--dwarf group sample), suggesting
that interactions may play a role in distributing the observed gas in
velocity space.

Second, the over-prediction of the low equivalent widths in the
$W_r(2796)-D$ superposition model (Section~\ref{sec:superboot}) may
also indicate a more complicated CGM in group environments than
assumed. In this scenario, the covering fraction around some
individual group galaxies may be less than expected in a superposition
model, which does account for the non-absorption present in the
isolated sample. Perhaps the ionization conditions or metallicities of
the gas are less consistently conducive to the presence of {\MgII}
absorption than in isolated galaxies even though multiple galaxies are
available to contribute absorbing material. However, the column
densities (which depend on ionization conditions, metallicities, and
path lengths) for the group sample are statistically comparable to the
isolated sample, suggesting that this is not the case. Alternatively,
and perhaps more simply, not every galaxy in the group contributes to
the absorption and the observed gas is more associated with an
intragroup medium rather than individual galaxies.

Third, the superposition model does not accurately represent the
absorption kinematics. A proper superposition that includes
galaxy--galaxy and absorber--galaxy velocity offsets results in a TPCF
with a velocity dispersion that is much larger than what is
observed. This is largely due to the galaxy--galaxy velocity offsets,
which we confine to $\Delta v \leq 500$~{\kms} in our group
definition. Therefore, the observed gas is likely coupled to the group
(intragroup medium) rather than individual member galaxies. Given that
the group environment kinematics are comparable to those associated
with isolated face-on galaxies probed along their minor axis
(presented in {\magiicat} V), this suggests that the intragroup
material is either outflowing material from one or more galaxies, or
is agitated similarly to outflows. This is also strengthened by the
fact that the average absorption spectrum for group environments has
larger optical depth at higher velocities than the average isolated
sample absorption spectrum. For a given line-of-sight velocity
$v\gtrsim 50$~{\kms}, group absorbers have either more gas, more
metal-rich gas, larger path lengths, or some combination compared to
isolated galaxies, but are similar in the cores of the absorption
profiles. If this is outflowing material, the fact that the gas
appears to be coupled to the group may suggest that it is gas
undergoing an ``intergalactic transfer'' by way of wind transfer as
described by \citet{angles17} in the FIRE simulations \citep[see
  also][]{oppenheimer08, keres09b, oppenheimer10}. In this scenario,
gas is transferred between nearby galaxies via outflowing winds and is
an accretion mode that dominates the accretion of gas onto $L^{\ast}$
galaxies by $z=0$.

Another possible explanation is that tidal stripping may agitate the
gas in similar ways. The \citet{angles17} simulations suggest that gas
stripping from galaxy interactions is less important than
intergalactic transfer except for the later stages of galaxy
mergers. We did not specifically target galaxies clearly undergoing
interactions and the later stages of mergers. However, warps and
potential tidal streams are directly observed in deep {\it HST} images
of at least one group in the sample \citep[Q1127$-$145, $z_{\rm
    abs}=0.313$;][]{ggk1127}. Also, because of the large radius of the
CGM in comparison to the visible portions of the host galaxy, we would
expect interactions to start changing CGM properties of the
participating galaxies before the visible galaxy portions become more
obvious. Thus we cannot rule out gas stripping and streams as the
source of {\MgII} absorption in groups.

There are further suggestions that merger/interaction activity is
giving rise to the observed group absorption. We examined the
properties of absorbers associated with galaxy--galaxy groups (i.e.,
the two brightest galaxies in a group have similar luminosities,
$L_1/L_2<3.5$) and those in galaxy--dwarf groups
($L_1/L_2\geq3.5$). Comparing the two, we found that absorbers in
galaxy--galaxy groups may have larger velocity dispersions and
equivalent widths than in galaxy--dwarf groups, although the result is
only marginally significant. The covering fractions are consistent
within uncertainties, with galaxy--galaxy groups trending toward
larger fractions. This result suggests that not only do galaxy--galaxy
interactions affect the CGM, but the type of interaction/environment
may influence the absorption properties. Groups in which major mergers
occur (galaxy--galaxy groups) may be more likely to cause tidal
stripping of CGM gas and/or induce star formation in both galaxies
involved. In the most dense environments of clusters, \citet{lopez08}
found an overabundance of strong {\MgII} absorbers whereas weak
absorbers are destroyed \citep[see also][]{padilla09,
  andrews13}. Combining our results with more dense environments, we
suggest that the group environment may enhance the absorption
strengths and kinematics, but once the environment becomes too dense,
and therefore too hot, this effect is reduced and the weakest
absorbers are eventually ionized to higher states. Further work is
needed to investigate this turn-over point for {\MgII}.

It is interesting that the group sample has only three non-absorbers,
with the rest having measurable absorption. This results in a covering
fraction of $f_c=0.89_{-0.09}^{+0.05}$ for the group environment
sample, in contrast to $f_c=0.68_{-0.03}^{+0.03}$ for the isolated
galaxy sample. If the superposition model is correct in that multiple
galaxies contribute to the observed absorption, then a larger covering
fraction in group environments would be expected. In our superposition
modeling, we in fact found a superposition covering fraction of
$f_c=0.83_{-0.01}^{+0.03}$ for the group environments, which is
consistent within uncertainties with the observed group environment
covering fraction. More importantly, this covering fraction is
significantly larger than that found in the isolated sample. Despite
the superposition model matching the observed group covering fraction,
it still does not accurately represent the observed kinematics. These
results combined further point to an intragroup medium for these group
environments (or more accurately, galaxy pairs in most cases), where
tidal stripping and intergalactic transfer is common for populating
the CGM with low-ionization, kinematically complex gas.

A potential bias in comparing the group and isolated environment
samples for the kinematics analysis is that the galaxies in the group
environment sample are located at a lower redshift on average than the
isolated sample: 0.411 versus 0.656, respectively, for the kinematics
sample only. However, in {\magiicat} IV we found that the kinematics
are consistent for blue galaxies at low and high redshift (split by
$\langle z_{\rm gal} \rangle = 0.656$) and the velocity dispersion
decreases from high to low redshift for red galaxies. If this redshift
bias were affecting the present analysis, the TPCFs for the group
environment sample would either remain constant or be more narrow than
the isolated sample. This is not the result we find; the gas
kinematics in the group environment sample are comparable to or more
active than for the isolated sample. As stated in
Section~\ref{sec:EWD}, a KS test comparing galaxy properties (impact
parameters, luminosities, colors, and redshifts) between the two
samples indicates that the null hypothesis that they were drawn from
the same population cannot be ruled out, so the galaxies themselves do
not appear to be different between samples with the information we
have available.

We have left out the sample of ultrastrong {\MgII} absorbers
associated with group environments found by \citet{nestor11} and
\citet{gauthier13} because they are outliers in equivalent width and
because we do not have their spectra. Additionally, the Nestor
{\etal}~absorbers were identified in low-resolution SDSS spectra, in
contrast to the high-resolution HIRES and UVES spectra for the sample
presented here. If these absorbers were included in the sample, the
mean equivalent widths, absorber velocity dispersions, covering
fractions, median column densities, and number of clouds would all
increase, in some cases making the group environment sample no longer
consistent with isolated galaxies. For example, if we include only the
\citet{gauthier13} absorber (4.2~{\AA}) in the kinematics analysis,
the resulting TPCF would be significantly more extended out to $\sim
550$~{\kms}. However, we do not include these absorbers in the sample
because they are extreme outliers in every absorption property. It is
possible that these ultrastrong {\MgII} absorbers are more likely
hosted by group environments due to their unique physical processes --
out of the isolated {\magiicat} sample of $\sim180$ galaxies, only one
is an ultrastrong absorber. Previous work has attributed these
absorbers to starburst-driven outflows from interactions and/or from
stripped material in the intragroup medium. However, further work
needs to be done with these absorbers to better understand their
origin.

Finally, the behavior of the low ionization {\MgII} doublet in group
environments differs from that of the intermediate, {\CIV}, and
higher, {\OVI}, ions. Recently, \citet{pointon17} showed that {\OVI}
associated with group galaxies similar to those presented here has
lower equivalent widths and a more narrow TPCF than around isolated
galaxies. Also, the covering fraction of {\OVI} in groups is less than
{\MgII} groups. The authors suggested that, similar to the results in
the EAGLE simulations by \citet{oppenheimer16}, {\OVI} is more
sensitive to the virial temperature and therefore the ionization
conditions of the host halo. Since group galaxies are hosted by more
massive halos, the absorbing gas is ionized to higher ionization
states, resulting in less observed {\OVI} absorption. A similar result
was found with {\CIV} by \citet{burchett16} at $z<0.015$, where the
detection rate for {\CIV} drops to zero when there are more than seven
galaxies in the group environment \citep[for cluster environments,
  see][]{burchett18}. They also found that the column densities appear
to be influenced by their host mass/environment, similar to the {\OVI}
and Oppenheimer {\etal}~work, but that {\CIV} may continue to be
observed in overdense regions due to containing more gas from
galaxy--galaxy interactions. In comparison, we have shown that {\MgII}
in groups ($2-5$ galaxies) may have larger covering fractions and
equivalent widths, and more optical depth at large line-of-sight
velocities compared to absorbers around isolated galaxies. This
suggests that {\MgII} may be less sensitive to the ionization
conditions of the host halo than the higher ionization states. Upon
reaching cluster sizes, {\MgII} halos are truncated and only the
weakest absorbers are destroyed \citep{lopez08, padilla09,
  andrews13}. This further suggests that the low and intermediate/high
ions trace different components of the CGM \citep[e.g.,][]{werk13,
  werk16, ford14, churchill15, muzahid15, stern16, nielsen17,
  pointon17} and emphasizes that a multiphase approach to studying the
CGM is necessary to fully understand the dominant mechanisms involved.

\section{Summary and Conclusions}
\label{sec:conclusions}

We presented the {\MgII} Absorber--Galaxy Catalog ({\magiicat}) group
sample to complement the isolated sample presented in our {\magiicat}
papers \citep{magiicat2, magiicat1, magiicat5, magiicat4,
  magiicat3}. The group sample consists of 29 {\MgII} absorbers
associated with group environments along 27 quasar sightlines for a
total of 74 foreground galaxies. The sample is located at $0.113 <
z_{\rm gal} < 0.888$ and within $D=200$~kpc of a background quasar
sightline. A group is defined as having two or more galaxies within a
projected distance of 200~kpc and with a velocity separation of less
than 500~{\kms}. With this sample, we examined the absorption
properties as a function of galaxy environment and find the following:

\begin{enumerate}[nolistsep]
  \item The median equivalent widths for the group environment sample
    ($0.65\pm0.13$~{\AA}) are larger than for isolated galaxies
    ($0.41\pm0.06$~{\AA}) ($1.7\sigma$).
  
  \item The equivalent width vs impact parameter anti-correlation may
    be flatter for galaxies in group environments than those in
    isolated environments, where a rank correlation test is marginally
    significant for the group environment sample at $2.9\sigma$
    compared to $7.9\sigma$ for isolated galaxies. If we assign the
    most luminous galaxy in the group as the absorber host, then the
    slope of the $W_r(2796)-D$ fit is significantly flatter than for
    isolated galaxies. The slopes are consistent within uncertainties
    when the group galaxy nearest to the quasar sightline is assumed
    to host the observed absorption.

  \item The covering fraction of {\MgII} in group environments,
    $f_c=0.89_{-0.09}^{+0.05}$, are larger than for isolated galaxies,
    $f_c=0.68_{-0.03}^{+0.03}$, although this is marginally
    significant at the $2.2\sigma$ level.
    
  \item Using the pixel-velocity TPCF method to study absorber
    kinematics, we found that while the velocity dispersion of
    absorbers in group environments is consistent within uncertainties
    compared to those in isolated environments ($0.8\sigma$), the
    group kinematics trend towards larger dispersions with more power
    at $\Delta v_{\rm pixel}=200$~{\kms}.

  \item The type of merger activity may influence the CGM
    properties. Groups in which the two brightest galaxies have
    similar luminosities (galaxy--galaxy; $L_1/L_2<3.5$) have
    $1.7\sigma$ ($1.8\sigma$) larger median (median) equivalent widths
    and larger absorber velocity dispersions ($2.5\sigma$) than in
    galaxy--dwarf groups ($L_1/L_2\geq3.5$). However, their covering
    fractions are comparable within uncertainties, with
    $f_c=0.95_{-0.11}^{+0.04}$ for galaxy--galaxy groups and
    $f_c=0.78_{-0.22}^{+0.14}$ for galaxy--dwarf groups.

  \item The distribution of fitted cloud column densities are
    consistent within uncertainties between the group and isolated
    samples. Absorbers in the group sample have a comparable number of
    clouds but a significantly ($3.3\sigma$) larger fraction of high
    velocity clouds, $v\geq 100$~{\kms}, than for the isolated
    sample. When only galaxy--galaxy group environments are compared
    to the isolated sample, the fraction of high velocity clouds in
    groups is increased.

  \item A superposition of individual group galaxy CGM results in
    equivalent widths that are comparable to the measured values in
    the group sample for the strongest absorbers. The model also finds
    a covering fraction of $f_c=0.83_{-0.01}^{+0.03}$, which is
    similar to the observed values. However, the superposition model
    is too simplistic to explain the observed TPCF (kinematic)
    distributions, where a proper superposition results in absorption
    velocity dispersions that are much too large.

  \item The group absorber kinematics appear similar to the kinematics
    of presumably outflowing gas around face-on galaxies probed along
    their minor axis (see {\magiicat} V). This suggests that the gas
    in group environments may be agitated similarly to that entrained
    in outflowing winds in isolated galaxies.

  \item We argue that the evidence presented here supports a model
    where the absorption associated with group environments forms an
    intragroup medium in which one or more galaxies contribute
    material, and where galaxy interactions distribute the gas
    throughout the group halo. The gas may be dispersed by outflows
    from one galaxy entering the intragroup medium and eventually
    falling onto another group member galaxy (intergalactic transfer)
    and/or by tidal stripping from interactions that remove gas from
    one galaxy and place it in the intragroup medium.

  \item Comparing our results to {\CIV} and {\OVI} in group
    environments, we find that the low and higher ions behave
    differently compared to their respective isolated samples,
    presenting further evidence that these ions trace different
    components within the CGM and intragroup medium.

\end{enumerate}

To better understand the gas traced by {\MgII} absorption, it would be
helpful to examine the kinematics of the gas relative to the
galaxy. While we have shown that absorbers associated with group
galaxies have larger velocity dispersions, we do not yet know if the
gas is being stripped from galaxies, accreting, or if the gas is truly
associated with a single galaxy or not. We have statistically shown
that the absorption is likely coupled to the group in an intragroup
medium rather than individual galaxies, but the complexity of galaxy
interactions may mean this is not always the case. More accurate
galaxy redshifts and rotation curves, estimates of galaxy star
formation rates, and deep surface brightness, high spatial resolution
imaging of the galaxies in groups will improve the situation.

\acknowledgments

N.M.N.~thanks John O'Meara for providing several reduced quasar
spectra. This material is based on work supported by the National
Science Foundation under grant No.~1210200 (NSF East Asia and Pacific
Summer Institutes). N.M.N.,~G.G.K.,~and M.T.M.~acknowledge the support
of the Australian Research Council through a Discovery Project
DP170103470. C.W.C.~acknowledges support by the National Science
Foundation under Grant No.~AST-1517816. S.K.P.~acknowledges support
through the Australian Government Research Training Program
Scholarship. M.T.M.~thanks the Australian Research Council for
Discovery Project grant DP130100568 which supported this
work. Observations for the Q1038$+$064 were obtained with the Apache
Point Observatory 3.5-meter telescope, which is owned and operated by
the Astrophysical Research Consortium. Some of the data presented
herein were obtained at the W. M. Keck Observatory, which is operated
as a scientific partnership among the California Institute of
Technology, the University of California and the National Aeronautics
and Space Administration. The Observatory was made possible by the
generous financial support of the W. M. Keck Foundation. Observations
were supported by Swinburne Keck program 2017A\_W248. The authors wish
to recognize and acknowledge the very significant cultural role and
reverence that the summit of Maunakea has always had within the
indigenous Hawaiian community.  We are most fortunate to have the
opportunity to conduct observations from this mountain.

{\it Facilities:} \facility{Keck:II (ESI)}, \facility{APO (DIS)}

\bibliographystyle{apj}
\bibliography{refs}

\clearpage

\addtocounter{table}{-3}

\begin{turnpage}
\begin{deluxetable*}{lllrrrrcccrcccrcc}
\tabletypesize{\footnotesize}
\tablecolumns{18}
\tablewidth{0pt}
\setlength{\tabcolsep}{0.06in}
\tablecaption{Observed Galaxy Properties \label{tab:obsprops}}\\
\tablehead{
  \colhead{}                         &
  \colhead{}                         &
  \colhead{}                         &
  \multicolumn{4}{c}{Galaxy ID}      &
  \colhead{}                         &
  \multicolumn{3}{c}{$B$-band}       &
  \colhead{}                         &
  \multicolumn{3}{c}{$K$-band}       &
  \colhead{}                         &
  \colhead{} \\
  \cline{4-7} \cline{9-11} \cline{13-15} \\[-5pt]
  \colhead{(1)}                      &
  \colhead{(2)}                      &
  \colhead{(3)}                      &
  \colhead{(4)}                      &
  \colhead{(5)}                      &
  \colhead{(6)}                      &
  \colhead{(7)}                      &
  \colhead{}                         &
  \colhead{(8)}                      &
  \colhead{(9)}                      &
  \colhead{(10)}                     &
  \colhead{}                         &
  \colhead{(11)}                     &
  \colhead{(12)}                     &
  \colhead{(13)}                     &
  \colhead{}                         &
  \colhead{(14)}                     \\
  \colhead{QSO~\tablenotemark{a}}    &
  \colhead{J-Name~\tablenotemark{a}} &
  \colhead{$z_{\rm gal}$}              &
  \colhead{$\Delta\alpha$}           &
  \colhead{$\Delta\delta$}           &
  \colhead{$\theta$}                 &
  \colhead{Ref~\tablenotemark{b}}    &
  \colhead{}                         &
  \colhead{$m_{y}$\tablenotemark{c}}  &
  \colhead{Band~\tablenotemark{d}}   &
  \colhead{Ref~\tablenotemark{b}}    &
  \colhead{}                         &
  \colhead{$m_{y}$\tablenotemark{e}}  &
  \colhead{Band~\tablenotemark{d}}   &
  \colhead{Ref~\tablenotemark{b}}    &
  \colhead{}                         &
  \colhead{SED~\tablenotemark{f}}    \\
  \colhead{}                         &
  \colhead{}                         &
  \colhead{}                         &
  \colhead{(arcsec)}                 &
  \colhead{(arcsec)}                 &
  \colhead{(arcsec)}                 &
  \multicolumn{11}{}{}               }
\startdata
\multirow{2}{*}{SDSS}                               & \multirow{2}{*}{J003340.21$-$005525.53}       & $0.1760  $ &   $-8.2$ &    $7.2$ & $ 10.91$ & 6  && $20.98$ & $g$(AB)                      & 14  &&  $19.72$ & $r$(AB)  & $14     $ && E/S0     \\[2pt]
                                                    &                                               & $0.1758  $ &   $18.9$ &    $2.7$ & $ 19.09$ & 6  && $21.01$ & $g$(AB)                      & 14  &&  $20.61$ & $r$(AB)  & $14     $ && Im       \\[2pt]\hline

\rule{0pt}{2.5ex}\multirow{2}{*}{SDSS}              & \multirow{2}{*}{J005244.23$-$005721.7}        & $0.13429 $ &   $-4.7$ &   $12.6$ & $ 13.42$ & 10 && $16.84$ & $r$(AB)                      & 14  &&  $13.53$ & $K_s$(V) & $15     $ && E/S0     \\[2pt]
                                                    &                                               & $0.13465 $ &   $-3.1$ &   $35.4$ & $ 35.55$ & 10 && $19.52$ & $g$(AB)                      & 14  &&  $19.52$ & $r$(AB)  & $14     $ && Scd      \\[2pt]\hline

\rule{0pt}{2.5ex}\multirow{5}{*}[-0.5em]{\bf 0150$-$202\tablenotemark{k}} & \multirow{5}{*}[-0.5em]{\bf J015227.32$-$200107.10} & $0.38260 $ & $8.5$ & $-7.8$ & $ 11.47$ & 3, 19\tablenotemark{h} && $21.15$ & F702W(V) & 3 && $\cdots$ & $\cdots$ & $\cdots $ && (Sbc)    \\[2pt]
                                                    &                                               & $0.38024 $ &   $-0.3$ &   $16.1$ & $ 16.10$ & 19 && $22.81$ & F702W(V)                     & 25  && $\cdots$ & $\cdots$ & $\cdots $ && (Sbc)    \\[2pt]
                                                    &                                               & $0.38146 $ &   $-7.8$ &   $15.9$ & $ 17.51$ & 19 && $22.63$ & F702W(V)                     & 25  && $\cdots$ & $\cdots$ & $\cdots $ && (Sbc)    \\
\enddata

\end{deluxetable*}
\end{turnpage}

\addtocounter{table}{-1}

\begin{turnpage}
\begin{deluxetable*}{lllrrrrcccrcccrcc}
\tabletypesize{\footnotesize}
\tablecolumns{18}
\tablewidth{0pt}
\setlength{\tabcolsep}{0.06in}
\tablecaption{(continued)}\\
\tablehead{
  \colhead{}                         &
  \colhead{}                         &
  \colhead{}                         &
  \multicolumn{4}{c}{Galaxy ID}      &
  \colhead{}                         &
  \multicolumn{3}{c}{$B$-band}       &
  \colhead{}                         &
  \multicolumn{3}{c}{$K$-band}       &
  \colhead{}                         &
  \colhead{} \\
  \cline{4-7} \cline{9-11} \cline{13-15} \\[-5pt]
  \colhead{(1)}                      &
  \colhead{(2)}                      &
  \colhead{(3)}                      &
  \colhead{(4)}                      &
  \colhead{(5)}                      &
  \colhead{(6)}                      &
  \colhead{(7)}                      &
  \colhead{}                         &
  \colhead{(8)}                      &
  \colhead{(9)}                      &
  \colhead{(10)}                     &
  \colhead{}                         &
  \colhead{(11)}                     &
  \colhead{(12)}                     &
  \colhead{(13)}                     &
  \colhead{}                         &
  \colhead{(14)}                     \\
  \colhead{QSO~\tablenotemark{a}}    &
  \colhead{J-Name~\tablenotemark{a}} &
  \colhead{$z_{\rm gal}$}              &
  \colhead{$\Delta\alpha$}           &
  \colhead{$\Delta\delta$}           &
  \colhead{$\theta$}                 &
  \colhead{Ref~\tablenotemark{b}}    &
  \colhead{}                         &
  \colhead{$m_{y}$\tablenotemark{c}}  &
  \colhead{Band~\tablenotemark{d}}   &
  \colhead{Ref~\tablenotemark{b}}    &
  \colhead{}                         &
  \colhead{$m_{y}$\tablenotemark{e}}  &
  \colhead{Band~\tablenotemark{d}}   &
  \colhead{Ref~\tablenotemark{b}}    &
  \colhead{}                         &
  \colhead{SED~\tablenotemark{f}}    \\
  \colhead{}                         &
  \colhead{}                         &
  \colhead{}                         &
  \colhead{(arcsec)}                 &
  \colhead{(arcsec)}                 &
  \colhead{(arcsec)}                 &
  \multicolumn{11}{}{}               }
\startdata
                 \multirow{2}{*}{\bf SDSS}          & \multirow{2}{*}{\bf J092554.71+400414.17}     & $0.2475  $ &   $-8.0$ &  $-20.8$ & $ 21.64$ & 20 && $20.28$ & $g$(AB)                      & 14  &&  $18.63$ & $r$(AB)  & $14     $ && E/S0     \\[2pt]
                                                    &                                               & $0.2467  $ &   $-7.2$ &  $-24.1$ & $ 24.69$ & 20 && $20.31$ & $g$(AB)                      & 14  &&  $19.55$ & $r$(AB)  & $14     $ && Sbc      \\[2pt]\hline

\rule{0pt}{2.5ex}\multirow{3}{*}[-0.2em]{\bf SDSS} & \multirow{3}{*}[-0.2em]{\bf J092837.98+602521.02} & $0.1537$ &  $-3.5$ &  $-14.7$ & $ 14.82$ & 20 && $20.64$ & $g$(AB)                      & 14  &&  $20.05$ & $r$(AB)  & $14     $ && Scd      \\[2pt]
                                                    &                                               & $0.1542  $ &   $30.2$ &  $-12.1$ & $ 19.19$ & 20 && $19.57$ & $g$(AB)                      & 14  &&  $18.99$ & $r$(AB)  & $14     $ && Scd      \\[2pt]
                                                    &                                               & $0.1540  $ &   $67.2$ &  $-12.3$ & $ 35.38$ & 20 && $19.47$ & $g$(AB)                      & 14  &&  $18.40$ & $r$(AB)  & $14     $ && E/S0     \\[2pt]\hline

\rule{0pt}{2.5ex}\multirow{2}{*}{\bf SDSS} & \multirow{2}{*}{\bf J100902.06+071343.87} & $0.35585$\tablenotemark{g} &    $3.2$ &   $0.03$ & $  3.13$ & 20 && $24.10$ & F390W(AB)                 & 24  &&  $23.21$ & F625W(AB)& $24     $ && Im       \\[2pt]
                                           &                                           & $0.35587$\tablenotemark{g} &    $1.7$ &   $-9.3$ & $  9.41$ & 20 && $21.65$ & $g$(AB)                   & 14  &&  $20.96$ & $r$(AB)  & $14     $ && Im       \\
\enddata

\end{deluxetable*}
\end{turnpage}

\addtocounter{table}{-1}

\begin{turnpage}
\begin{deluxetable*}{lllrrrrcccrcccrcc}
\tabletypesize{\footnotesize}
\tablecolumns{18}
\tablewidth{0pt}
\setlength{\tabcolsep}{0.06in}
\tablecaption{(continued)}\\
\tablehead{
  \colhead{}                         &
  \colhead{}                         &
  \colhead{}                         &
  \multicolumn{4}{c}{Galaxy ID}      &
  \colhead{}                         &
  \multicolumn{3}{c}{$B$-band}       &
  \colhead{}                         &
  \multicolumn{3}{c}{$K$-band}       &
  \colhead{}                         &
  \colhead{} \\
  \cline{4-7} \cline{9-11} \cline{13-15} \\[-5pt]
  \colhead{(1)}                      &
  \colhead{(2)}                      &
  \colhead{(3)}                      &
  \colhead{(4)}                      &
  \colhead{(5)}                      &
  \colhead{(6)}                      &
  \colhead{(7)}                      &
  \colhead{}                         &
  \colhead{(8)}                      &
  \colhead{(9)}                      &
  \colhead{(10)}                     &
  \colhead{}                         &
  \colhead{(11)}                     &
  \colhead{(12)}                     &
  \colhead{(13)}                     &
  \colhead{}                         &
  \colhead{(14)}                     \\
  \colhead{QSO~\tablenotemark{a}}    &
  \colhead{J-Name~\tablenotemark{a}} &
  \colhead{$z_{\rm gal}$}              &
  \colhead{$\Delta\alpha$}           &
  \colhead{$\Delta\delta$}           &
  \colhead{$\theta$}                 &
  \colhead{Ref~\tablenotemark{b}}    &
  \colhead{}                         &
  \colhead{$m_{y}$\tablenotemark{c}}  &
  \colhead{Band~\tablenotemark{d}}   &
  \colhead{Ref~\tablenotemark{b}}    &
  \colhead{}                         &
  \colhead{$m_{y}$\tablenotemark{e}}  &
  \colhead{Band~\tablenotemark{d}}   &
  \colhead{Ref~\tablenotemark{b}}    &
  \colhead{}                         &
  \colhead{SED~\tablenotemark{f}}    \\
  \colhead{}                         &
  \colhead{}                         &
  \colhead{}                         &
  \colhead{(arcsec)}                 &
  \colhead{(arcsec)}                 &
  \colhead{(arcsec)}                 &
  \multicolumn{11}{}{}               }
\startdata
                 \multirow{2}{*}{\bf 1623+269}      & \multirow{2}{*}{\bf J162548.79+264658.75}     & $0.888   $ &   $-1.0$ &    $6.1$ & $  6.21$ & 3  && $23.63$ & F702W(V)                     & 3   &&  $18.30$ & $K_s$(V) & $8      $ && E/S0     \\[2pt]
                                                    &                                               & $0.888   $ &   $-2.8$ &    $8.8$ & $  9.27$ & 3  && $23.59$ & F702W(V)                     & 3   && $\cdots$ & $\cdots$ & $\cdots $ && (Sbc)    \\[2pt]\hline

\rule{0pt}{2.5ex}\multirow{2}{*}{SDSS}              & \multirow{2}{*}{J204431.46+011312.43}         & $0.1921  $ &    $6.1$ &   $-3.6$ & $  7.08$ & 6  && $21.40$ & $g$(AB)                      & 14  &&  $20.66$ & $r$(AB)  & $14     $ && Sbc      \\[2pt]
                                                    &                                               & $0.1927  $ &   $-2.1$ &   $-7.5$ & $  7.79$ & 6  && $20.15$ & $g$(AB)                      & 14  &&  $18.93$ & $r$(AB)  & $14     $ && E/S0     \\[2pt]\hline

\rule{0pt}{2.5ex}\multirow{4}{*}[-0.4em]{\bf 2126$-$158} & \multirow{4}{*}[-0.4em]{\bf J212912.17$-$153841.04} & $0.6668  $ &    $6.3$ &   $-3.7$ & $  7.10$ & 16 && $20.79$ & $i'(AB)$          & 16  && $\cdots$ & $\cdots$ & $\cdots $ && (Sbc)    \\[2pt]
                                                    &                                               & $0.6643  $ &    $8.3$ &   $-3.2$ & $  8.60$ & 16 && $20.34$ & $i'(AB)$                     & 16  && $\cdots$ & $\cdots$ & $\cdots $ && (Sbc)    \\[2pt]
                                                    &                                               & $0.6647  $ &   $12.6$ &   $-2.4$ & $ 12.40$ & 16 && $22.08$ & $i'(AB)$                     & 16  && $\cdots$ & $\cdots$ & $\cdots $ && (Sbc)    \\
\enddata

\tablenotetext{a}{Groups included in the kinematics analysis are
  marked with bold-faced field names. We have the HIRES/Keck or
  UVES/VLT spectra for each bolded group, and have measurable {\MgII}
  above our detection threshold.}

\tablenotetext{b}{Galaxy Identification and Apparent Magnitude
  Reference: (1) \citet{guillemin97}, (2) this work, 
  (3) \citet{kcems11}, (4) \citet{steidel97},
  (6) \citet{chen10a}, (7) \citet{sdp94}, (8) Steidel (personal
  communication), (9) \citet{ggk1127}, (10) \citet{kacprzak11kin},
  (14) NED/SDSS, (15) NED/2MASS, (16) \citet{whiting06}, (17)
  \citet{bielby17}, (18) \citet{peroux17}, (19) \citet{rahmani18},
  (20) \citet{werk12}, (21) \citet{chen09}, (22) \citet{chen01b}, (23)
  \citet{johnson13}, (24) \citet{meiring11}, and (25) this work.}

\tablenotetext{c}{Apparent magnitude used to obtain $M_{B}$.}
\tablenotetext{d}{Magnitude Band and Type: (AB) AB magnitude, and (V)
  Vega magnitude.}
\tablenotetext{e}{Apparent magnitude used to obtain $M_K$.}
\tablenotetext{f}{Galaxy Spectral Energy Distributions: (Sbc) No color
  information -- Sbc used.}
\tablenotetext{g}{Redshift measured from Keck/ESI spectrum (this work).}
\tablenotetext{h}{The right ascension and declination for this galaxy
  reported by \citet{rahmani18} is incorrect.}
\tablenotetext{k}{Originally included as an isolated galaxy in
  {\magiicat} \citep{magiicat1}.}

\end{deluxetable*}
\end{turnpage}


\begin{turnpage}
\begin{deluxetable*}{llllccrrrrccrrcc}
\tabletypesize{\footnotesize}
\tablecolumns{16}
\tablewidth{0pt}
\setlength{\tabcolsep}{0.04in}
\tablecaption{Calculated Galaxy and Absorption Properties \label{tab:calcprops}}
\tablehead{
  \colhead{}                         &
  \colhead{}                         &
  \colhead{}                         &
  \multicolumn{4}{c}{{\MgII} Absorption} &  
  \colhead{}                         &
  \multicolumn{3}{c}{$B$-band}       &
  \colhead{}                         &
  \multicolumn{3}{c}{$K$-band}       &
  \colhead{}                         \\
  \cline{4-7} \cline{9-11} \cline{13-15} \\[-5pt]
  \colhead{(1)}                      &
  \colhead{(2)}                      &
  \colhead{(3)}                      &
  \colhead{(4)}                      &
  \colhead{(5)}                      &
  \colhead{(6)}                      &
  \colhead{(7)}                      &
  \colhead{(8)}                      &
  \colhead{(9)}                      &
  \colhead{(10)}                     &
  \colhead{(11)}                     &
  \colhead{}                         &
  \colhead{(12)}                     &
  \colhead{(13)}                     &
  \colhead{(14)}                     &
  \colhead{(15)}                     \\
  \colhead{QSO~\tablenotemark{a}}    &
  \colhead{J-Name~\tablenotemark{a}} &
  \colhead{$z_{\rm gal}$}              &
  \colhead{$z_{\rm abs}$}              &
  \colhead{$W_r(2796)$}              &
  \colhead{$DR$}                     &
  \colhead{Ref~\tablenotemark{b}}    &
  \colhead{$D$}                      &
  \colhead{$K_{By}$\tablenotemark{c}} &
  \colhead{$M_B$\tablenotemark{d}}   &
  \colhead{$L_B/L_B^{\ast}$}          &
  \colhead{}                         &
  \colhead{$K_{Ky}$\tablenotemark{e}} &
  \colhead{$M_K$\tablenotemark{d}}   &
  \colhead{$L_K/L_K^{\ast}$}           &
  \colhead{$B-K$}                    \\
  \colhead{}                         &
  \colhead{}                         &
  \colhead{}                         &
  \colhead{}                         &
  \colhead{\AA}                      &
  \colhead{}                         &
  \colhead{}                         &
  \colhead{(kpc)}                    &
  \colhead{}                         &
  \colhead{}                         &
  \colhead{}                         &
  \colhead{}                         &
  \colhead{}                         &
  \colhead{}                         &
  \colhead{}                         &
  \colhead{}                         }    
\startdata
                 \multirow{2}{*}{SDSS}              & \multirow{2}{*}{J003340.21$-$005525.53}       & $0.1760  $ & \multirow{2}{*}{$0.1759  $} & \multirow{2}{*}{$0.19\pm0.04$}     & \multirow{2}{*}{$\cdots$}        & \multirow{2}{*}{6}  & $  32.3$ & $ 0.40$ & $-19.07$ & $0.23$ && $     0.31$ & $ -21.26$ & $  0.29$ & $2.18  $ \\[2pt]
                                                    &                                               & $0.1758  $ &                             &                                    &                                  &                     & $  56.4$ & $ 0.02$ & $-18.66$ & $0.16$ && $    -0.05$ & $ -19.27$ & $  0.04$ & $0.61  $ \\[2pt]\hline

\rule{0pt}{2.5ex}\multirow{2}{*}{SDSS}              & \multirow{2}{*}{J005244.23$-$005721.7}        & $0.13429 $ & \multirow{2}{*}{$0.1346  $} & \multirow{2}{*}{$1.46\pm0.04$}     & \multirow{2}{*}{$1.190\pm0.05$}  & \multirow{2}{*}{10} & $  31.7$ & $-0.94$ & $-21.22$ & $1.74$ && $    -0.26$ & $ -23.39$ & $  2.14$ & $2.16  $ \\[2pt]
                                                    &                                               & $0.13465 $ &                             &                                    &                                  &                     & $  84.3$ & $ 0.03$ & $-19.01$ & $0.23$ && $     0.05$ & $ -19.99$ & $  0.09$ & $0.97  $ \\[2pt]\hline

\rule{0pt}{2.5ex}\multirow{5}{*}[-0.5em]{\bf 0150$-$202\tablenotemark{k}} & \multirow{5}{*}[-0.5em]{\bf J015227.32$-$200107.10} & $0.38260$ & \multirow{5}{*}[-0.5em]{$0.383074$} & \multirow{5}{*}[-0.5em]{$0.168\pm0.015$} & \multirow{5}{*}[-0.5em]{$1.17\pm0.17$} & \multirow{5}{*}[-0.5em]{14} & $59.6$ & $-1.02$ & $-19.49$ & $0.27$ && $\cdots$ & $\cdots$ & $\cdots$ & $\cdots$ \\
\enddata

\end{deluxetable*}
\end{turnpage}

\addtocounter{table}{-1}

\begin{turnpage}
\begin{deluxetable*}{llllccrrrrccrrcc}
\tabletypesize{\footnotesize}
\tablecolumns{16}
\tablewidth{0pt}
\setlength{\tabcolsep}{0.04in}
\tablecaption{(continued)}
\tablehead{
  \colhead{}                         &
  \colhead{}                         &
  \colhead{}                         &
  \multicolumn{4}{c}{{\MgII} Absorption} &  
  \colhead{}                         &
  \multicolumn{3}{c}{$B$-band}       &
  \colhead{}                         &
  \multicolumn{3}{c}{$K$-band}       &
  \colhead{}                         \\
  \cline{4-7} \cline{9-11} \cline{13-15} \\[-5pt]
  \colhead{(1)}                      &
  \colhead{(2)}                      &
  \colhead{(3)}                      &
  \colhead{(4)}                      &
  \colhead{(5)}                      &
  \colhead{(6)}                      &
  \colhead{(7)}                      &
  \colhead{(8)}                      &
  \colhead{(9)}                      &
  \colhead{(10)}                     &
  \colhead{(11)}                     &
  \colhead{}                         &
  \colhead{(12)}                     &
  \colhead{(13)}                     &
  \colhead{(14)}                     &
  \colhead{(15)}                     \\
  \colhead{QSO~\tablenotemark{a}}    &
  \colhead{J-Name~\tablenotemark{a}} &
  \colhead{$z_{\rm gal}$}              &
  \colhead{$z_{\rm abs}$}              &
  \colhead{$W_r(2796)$}              &
  \colhead{$DR$}                     &
  \colhead{Ref~\tablenotemark{b}}    &
  \colhead{$D$}                      &
  \colhead{$K_{By}$\tablenotemark{c}} &
  \colhead{$M_B$\tablenotemark{d}}   &
  \colhead{$L_B/L_B^{\ast}$}          &
  \colhead{}                         &
  \colhead{$K_{Ky}$\tablenotemark{e}} &
  \colhead{$M_K$\tablenotemark{d}}   &
  \colhead{$L_K/L_K^{\ast}$}           &
  \colhead{$B-K$}                    \\
  \colhead{}                         &
  \colhead{}                         &
  \colhead{}                         &
  \colhead{}                         &
  \colhead{\AA}                      &
  \colhead{}                         &
  \colhead{}                         &
  \colhead{(kpc)}                    &
  \colhead{}                         &
  \colhead{}                         &
  \colhead{}                         &
  \colhead{}                         &
  \colhead{}                         &
  \colhead{}                         &
  \colhead{}                         &
  \colhead{}                         }    
\startdata
                 \multirow{2}{*}{\bf SDSS}          & \multirow{2}{*}{\bf J092554.71+400414.17}     & $0.2475  $ & \multirow{2}{*}{$0.247604$} & \multirow{2}{*}{$1.18\pm0.14$}     & \multirow{2}{*}{$1.23\pm0.22$}   & \multirow{2}{*}{14} & $  84.0$ & $ 0.76$ & $-21.25$ & $1.57$ && $     1.37$ & $ -23.21$ & $  1.67$ & $1.96  $ \\[2pt]
                                                    &                                               & $0.2467  $ &                             &                                    &                                  &                     & $  95.6$ & $ 0.36$ & $-20.52$ & $0.80$ && $     1.06$ & $ -21.98$ & $  0.54$ & $1.46  $ \\[2pt]\hline

\rule{0pt}{2.5ex}\multirow{3}{*}[-0.2em]{\bf SDSS} & \multirow{3}{*}[-0.2em]{\bf J092837.98+602521.02} & $0.1537$ & \multirow{3}{*}[-0.2em]{$0.153783$} & \multirow{3}{*}[-0.2em]{$1.16\pm0.16$} & \multirow{3}{*}[-0.2em]{$1.10\pm0.22$} & \multirow{3}{*}[-0.2em]{14} & $  39.5$ & $ 0.07$ & $-18.76$ & $0.18$ && $     0.47$ & $ -19.75$ & $  0.07$ & $0.99  $ \\[2pt]
                                                    &                                               & $0.1542  $ &                             &                                    &                                  &                     & $  51.3$ & $ 0.07$ & $-19.84$ & $0.48$ && $     0.47$ & $ -20.81$ & $  0.20$ & $0.97  $ \\[2pt]
                                                    &                                               & $0.1540  $ &                             &                                    &                                  &                     & $  94.5$ & $ 0.28$ & $-20.14$ & $0.63$ && $     1.23$ & $ -22.16$ & $  0.68$ & $2.02  $ \\
\enddata

\end{deluxetable*}
\end{turnpage}

\addtocounter{table}{-1}

\begin{turnpage}
\begin{deluxetable*}{llllccrrrrccrrcc}
\tabletypesize{\footnotesize}
\tablecolumns{16}
\tablewidth{0pt}
\setlength{\tabcolsep}{0.04in}
\tablecaption{(continued)}
\tablehead{
  \colhead{}                         &
  \colhead{}                         &
  \colhead{}                         &
  \multicolumn{4}{c}{{\MgII} Absorption} &  
  \colhead{}                         &
  \multicolumn{3}{c}{$B$-band}       &
  \colhead{}                         &
  \multicolumn{3}{c}{$K$-band}       &
  \colhead{}                         \\
  \cline{4-7} \cline{9-11} \cline{13-15} \\[-5pt]
  \colhead{(1)}                      &
  \colhead{(2)}                      &
  \colhead{(3)}                      &
  \colhead{(4)}                      &
  \colhead{(5)}                      &
  \colhead{(6)}                      &
  \colhead{(7)}                      &
  \colhead{(8)}                      &
  \colhead{(9)}                      &
  \colhead{(10)}                     &
  \colhead{(11)}                     &
  \colhead{}                         &
  \colhead{(12)}                     &
  \colhead{(13)}                     &
  \colhead{(14)}                     &
  \colhead{(15)}                     \\
  \colhead{QSO~\tablenotemark{a}}    &
  \colhead{J-Name~\tablenotemark{a}} &
  \colhead{$z_{\rm gal}$}              &
  \colhead{$z_{\rm abs}$}              &
  \colhead{$W_r(2796)$}              &
  \colhead{$DR$}                     &
  \colhead{Ref~\tablenotemark{b}}    &
  \colhead{$D$}                      &
  \colhead{$K_{By}$\tablenotemark{c}} &
  \colhead{$M_B$\tablenotemark{d}}   &
  \colhead{$L_B/L_B^{\ast}$}          &
  \colhead{}                         &
  \colhead{$K_{Ky}$\tablenotemark{e}} &
  \colhead{$M_K$\tablenotemark{d}}   &
  \colhead{$L_K/L_K^{\ast}$}           &
  \colhead{$B-K$}                    \\
  \colhead{}                         &
  \colhead{}                         &
  \colhead{}                         &
  \colhead{}                         &
  \colhead{\AA}                      &
  \colhead{}                         &
  \colhead{}                         &
  \colhead{(kpc)}                    &
  \colhead{}                         &
  \colhead{}                         &
  \colhead{}                         &
  \colhead{}                         &
  \colhead{}                         &
  \colhead{}                         &
  \colhead{}                         &
  \colhead{}                         }    
\startdata
                 \multirow{2}{*}{\bf 1623+269}      & \multirow{2}{*}{\bf J162548.79+264658.75}     & $0.888   $ & \multirow{2}{*}{$0.887679$} & \multirow{2}{*}{$0.903\pm0.004$}   & \multirow{2}{*}{$1.245\pm0.01$}  & \multirow{2}{*}{3}  & $  47.9$ & $ 0.12$ & $-20.36$ & $0.34$ && $    -0.64$ & $ -23.01$ & $  1.02$ & $2.65  $ \\[2pt]
                                                    &                                               & $0.888   $ &                             &                                    &                                  &                     & $  71.4$ & $-0.28$ & $-19.99$ & $0.24$ && $\cdots   $ & $ \cdots$ & $\cdots$ & $\cdots$ \\[2pt]\hline

\rule{0pt}{2.5ex}\multirow{2}{*}{SDSS}              & \multirow{2}{*}{J204431.46+011312.43}         & $0.1921  $ & \multirow{2}{*}{$0.1927  $} & \multirow{2}{*}{$0.50\pm0.08$}     & \multirow{2}{*}{$\cdots$}        & \multirow{2}{*}{6}  & $  22.5$ & $ 0.21$ & $-18.67$ & $0.15$ && $     0.16$ & $ -19.98$ & $  0.08$ & $1.31  $ \\[2pt]
                                                    &                                               & $0.1927  $ &                             &                                    &                                  &                     & $  24.8$ & $ 0.49$ & $-20.20$ & $0.64$ && $     0.33$ & $ -22.20$ & $  0.68$ & $2.00  $ \\[2pt]\hline

\rule{0pt}{2.5ex}\multirow{4}{*}[-0.4em]{\bf 2126$-$158} & \multirow{4}{*}[-0.4em]{\bf J212912.17$-$153841.04} & $0.6668$ & \multirow{4}{*}[-0.4em]{$0.662742$} & \multirow{4}{*}[-0.4em]{$1.903\pm0.014$} & \multirow{4}{*}[-0.4em]{$1.14\pm0.02$} & \multirow{4}{*}[-0.4em]{14} & $  49.7$ & $-0.74$ & $-21.49$ & $1.22$ && $\cdots   $ & $ \cdots$ & $\cdots$ & $\cdots$ \\
\enddata

\tablenotetext{a}{Groups included in the kinematics analysis are
  marked with bold-faced field names. We have the HIRES/Keck or
  UVES/VLT spectra for each bolded group, and have measurable {\MgII}
  above our detection threshold.}

\tablenotetext{b}{{\MgII} Absorption Measurements: (1)
  \citet{guillemin97},
  (3) \citet{kcems11},
  (6) \citet{chen10a},
  (10) \citet{kacprzak11kin},
  (13) \citet{evans-thesis}, and (14) This work.}

\tablenotetext{c}{$K$-correction used to obtain $M_B$ from column (8)
  in Table 1 -- Observed Galaxy Properties.}
\tablenotetext{d}{Absolute magnitudes are AB magnitudes.}
\tablenotetext{e}{$K$-correction used to obtain $M_K$ from column (11)
  in Table 1 -- Observed Galaxy Properties.}

\tablenotetext{f}{$R$-band absolute magnitude, $M_R$, and luminosity,
  $L_R/L_R^{\ast}$, obtained from \citet{peroux17}.}

\tablenotetext{g}{Redshift measured from Keck/ESI spectrum (this work).}

\tablenotetext{k}{Originally included as an isolated galaxy in
  {\magiicat} \citep{magiicat1}.}

\end{deluxetable*}
\end{turnpage}

\end{document}